\documentclass[journal,12pt,onecolumn,draftclsnofoot]{IEEEtran}
\usepackage{algpseudocode}
\usepackage{pict2e}
\usepackage{cite}
\usepackage{supertabular}
\usepackage{booktabs}
\usepackage{tikz}

\usetikzlibrary{calc, shadings, shadows, shapes.arrows}
\usetikzlibrary{decorations.pathreplacing}
\usetikzlibrary{shapes}
\usepackage{amsmath}
\usepackage{nccmath}
\usepackage{amssymb}
\usepackage{amsthm}
\usepackage{bbm}
\usepackage[algoruled, linesnumbered]{algorithm2e}
\usepackage{algpseudocode}
\usepackage{graphicx}
\usepackage{caption}
\usepackage{subcaption}
\usepackage{verbatim}
\usepackage{longtable}
\usepackage{array}
\usepackage{arydshln}
\usepackage{lscape}
\usepackage{multirow}
\usepackage{makecell}
\usepackage{slashbox}	
\usepackage{enumitem}

\setlist[itemize]{leftmargin=*}
\graphicspath{{Figs/}}
\theoremstyle{plain} \newtheorem{thm}{Theorem}
\theoremstyle{plain} 
\theoremstyle{definition} 
\theoremstyle{definition} 
\theoremstyle{definition} 

\makeatletter
\def\bbordermatrix#1{\begingroup \m@th
	\@tempdima 4.75\p@
	\setbox\z@\vbox{%
		\def\cr{\crcr\noalign{\kern2\p@\global\let\cr\endline}}%
		\ialign{$##$\hfil\kern2\p@\kern\@tempdima&\thinspace\hfil$##$\hfil
			&&\quad\hfil$##$\hfil\crcr
			\omit\strut\hfil\crcr\noalign{\kern-\baselineskip}%
			#1\crcr\omit\strut\cr}}%
	\setbox\tw@\vbox{\unvcopy\z@\global\setbox\@ne\lastbox}%
	\setbox\tw@\hbox{\unhbox\@ne\unskip\global\setbox\@ne\lastbox}%
	\setbox\tw@\hbox{$\kern\wd\@ne\kern-\@tempdima\left[\kern-\wd\@ne
		\global\setbox\@ne\vbox{\box\@ne\kern2\p@}%
		\vcenter{\kern-\ht\@ne\unvbox\z@\kern-\baselineskip}\,\right]$}%
	\null\;\vbox{\kern\ht\@ne\box\tw@}\endgroup}
\makeatother

\title{\huge{Curriculum Learning for Goal-Oriented Semantic Communications with a Common Language}}

\author{\small{Mohammad Karimzadeh Farshbafan$^\dagger$, Walid Saad$^\dagger$, and Merouane Debbah$^+$} \\
{\footnotesize{$^\dagger$ Wireless@VT, Bradley Department of Electrical and Computer Engineering, Virginia Tech, Blacksburg, VA, USA, \\
$^+$ Technology Innovation Institute, Abu Dhabi, United Arab Emirates, and \\
Mohamed Bin Zayed University of Artificial Intelligence, 9639 Masdar City, Abu Dhabi, United Arab Emirates, \\
emails: $^\dagger$ $\{$mkarimzadeh, walids$\}$@vt.edu, $^+$ $\text{merouane.debbah}$@tii.ae}}}

\allowdisplaybreaks[1]

\begin{document}

	\maketitle
\vspace{-20mm}	
\begin{abstract}
Goal-oriented semantic communication will be a pillar of next-generation wireless networks. Despite significant recent efforts in this area, most prior works are focused on specific data types (e.g., image or audio), and they ignore the goal and effectiveness aspects of semantic transmissions. In contrast, in this paper, a holistic goal-oriented semantic communication framework is proposed to enable a speaker and a listener to cooperatively execute a set of sequential tasks in a dynamic environment. A common language based on a hierarchical belief set is proposed to enable semantic communications between speaker and listener. The speaker, acting as an observer of the environment, utilizes the beliefs to transmit an initial description of its observation (called event) to the listener. The listener is then able to infer on the transmitted description and complete it by adding related beliefs to the transmitted beliefs of the speaker. As such, the listener reconstructs the observed event based on the completed description, and it then takes appropriate action in the environment based on the reconstructed event. An optimization problem is defined to determine the perfect and abstract description of the events while minimizing the various communication costs with constraints on the task execution time and belief efficiency. Then, a novel bottom-up curriculum learning (CL) framework based on reinforcement learning is proposed to solve the optimization problem and enable the speaker and listener to gradually identify the structure of the belief set and the perfect and abstract description of the events. Simulation results show that the proposed CL method outperforms classical RL and CL without inference scheme in terms of convergence time, task execution cost and time, reliability, and belief efficiency.
\end{abstract}
\vspace{-6mm}
\begin{IEEEkeywords}
\vspace{-3mm}
	Goal-Oriented Semantic Communication, Semantic Optimization, Curriculum Learning, Reinforcement Learning.
\end{IEEEkeywords}
\vspace{-6mm}
\section{Introduction}
\vspace{-1mm}
Goal-oriented communication will be a pillar of next-generation wireless networks due to the radical increase in the autonomy level of emerging communication services in the Internet of Everything (IoE) \cite{popovski2020semantic, saad2019vision, chaccour2022less}.
For example, remotely controlling high-precision manufacturing in automated factories can be a meaningful application of goal-oriented communication.
However, maximizing the bit accuracy, which is the main goal of current networks, is not sufficient for goal-oriented communication. 
Goal-oriented communication requires the network to consider the \emph{semantics (meanings) and effectiveness} \cite{popovski2020semantic} of the data transmission.
Hence, current metrics and algorithms for optimizing wireless networks must be redesigned to consider semantics and effectiveness. 
Deploying semantic communications for goal-oriented networks requires overcoming many challenges \cite{strinati20216g, luo2022semantic} such as modeling semantic information, defining semantic and effectiveness metrics, and enabling transmitter-receiver cooperation for goal-oriented communications.
\vspace{-14mm}
\subsection{Prior Works}
\vspace{-2mm}
Several recent works investigated the principles of semantic communications \cite{zhou2022cognitive, liu2022indirect, yang2022semantic}. In \cite{zhou2022cognitive} and \cite{liu2022indirect}, the authors studied the information-theoric challenges of semantic communications.
In \cite{zhou2022cognitive}, the authors proposed a cognitive semantic communication system using a knowledge graph for semantic extraction (SE) and enhanced error correction. The work in \cite{liu2022indirect} investigated the rate-distortion function for a semantic source model, which is not observable for the transmitter and can only be inferred from an extrinsic observation. 
In \cite{yang2022semantic}, the authors studied the implementation overhead of training SE models by introducing a solution based on federated learning (FL) and edge-based knowledge graphs. Despite being interesting, the works in \cite{zhou2022cognitive, liu2022indirect, yang2022semantic} did not consider the effectiveness of the transmitted information for goal-oriented communications.

The works in \cite{zhou2021semantic, farsad2018deep, lu2021reinforcement} utilized semantic communication to improve the transmission of specific data types (e.g., text, image, and audio). The authors in \cite{zhou2021semantic} introduced a rigorous semantic communication system based on a universal adaptive transformer for text transmission.
In \cite{farsad2018deep}, the authors proposed a semantic communication framework for text transmission using deep learning (DL). The work in \cite{lu2021reinforcement} introduced a reinforcement learning (RL) solution to capture the meanings of transmitted information by learning semantic similarities. 
The main drawback of \cite{zhou2021semantic, farsad2018deep, lu2021reinforcement} is that their models are only applicable to specific data types and, like \cite{zhou2022cognitive, liu2022indirect, yang2022semantic}, they do not consider the effectiveness of the semantics on the system's goal and, thus, they cannot be generalized to broader goal-oriented semantic communication scenarios.

{\color{black}The work in \cite{du2022rethinking} investigated the security aspect of semantic communication for Internet-of-Things (IoT) and introduced two novel metrics named semantic secrecy outage probability and detection failure probability for the security of semantic-based IoT systems.
The authors in \cite{zhang2022unified} proposed a unified task-based semantic communication system using a DL method, which simultaneously supports tasks with different data modalities, including image, text, and speech.} The work in \cite{yun2021attention} introduced a model for implementing semantic communications to address the reliability and latency requirements for drone networks. 
In \cite{lotfi2021semantic}, the authors proposed a semantic-aware collaborative deep RL (DRL) to select the best subset of semantically relevant heterogeneous DRL agents across a wireless cellular network.
The works in \cite{yun2021attention, lotfi2021semantic} can be considered as task-oriented semantic communication approaches. However, their solutions are only applicable to the specific applications that they considered and, thus, they cannot be generalized. Meanwhile, the authors in \cite{seo2021semantics} proposed a semantic-native communication structure to extract the most effective semantics of the transmitter for the receiver. However, their approach cannot capture the effectiveness of the transmitted semantics
for task execution purposes.

\vspace{-6mm}
\subsection{Contributions}
\vspace{-2mm}
The main contribution of this paper is a novel semantic communication system that leverages speaker and listener cooperation to perform goal-oriented communications while optimizing task execution time and cost, reliability, and belief efficiency. Our main contributions include:

{\color{black}
  \begin{itemize}
    \item We introduce a holistic model for goal-oriented communications, which includes a speaker and a listener who wish to jointly execute a set of tasks in a dynamic environment. Each task is defined as a chain of multiple sequential events named task event chain (TEC), whereby each event observed by the speaker captures the state of the environment at each time. 
    The speaker must inform the listener about the observed events, and the listener is responsible for taking a proper action regarding the observed event to steer the system to its goal.
    \item For event description and the communication between the speaker and listener, we assume the existence of a basic \emph{common language set} called \emph{belief set} with a hierarchical structure, in which the beliefs can be categorized in different levels, based on the amount of semantic information that they can convey and their data type. For event description, the speaker transmits an abstract description of the observed event using the belief set. Then, the listener infers and completes the description based on the received information.
    \item We formulate an optimization problem to minimize the costs incurred to speaker and listener with constraints on the task execution time and belief efficiency. To efficiently solve this problem, we develop a novel bottom-up curriculum learning (CL) \cite{JMLR:v21:20-212} framework based on RL. The main objective of the proposed CL is to gradually identifying the hierarchical structure of the belief set and the perfect description of the events. We analytically derive sufficient conditions under which the proposed CL method can find a perfect description of each event.
    \item Simulation results show that our CL method outperforms classical RL and CL without inference in all metrics. In particular, our CL method compared to classical RL and CL without inference yields around $89\%$ and $65.6\%$ reduction in the task execution time, and $94\%$ and $80.3\%$ reduction in the task execution cost, $45\%$ and $1.2\%$ improvement in task execution reliability, and $2.5$-fold and $15.5\%$ improvement in belief efficiency.
   
\end{itemize}
}

The rest of the paper is organized as follows. The proposed goal-oriented semantic communication system is presented in Section \ref{Section:System model}. Then, we present the proposed CL framework in Section \ref{Section:Proposed_Method}. Simulation results are analyzed in Section \ref{Section:Numerical_Results}, and conclusions are drawn in Section \ref{Section:Conclusion}.
\vspace{-5mm}
\section{System Model}
\vspace{-2mm}
\label{Section:System model}
Consider a speaker and listener who want to cooperatively execute a set of sequential tasks in an environment observed by the speaker. Each task consists of a chain of sequential events called TEC. At a given time, an event occurs in the environment and is observed by the speaker. This observed event captures the state of the environment, as perceived by the speaker at the observation time. 
The distortion level in the observed event of the speaker is relatively tolerable such that the observed event does not misinform about the state of the environment \cite{seo2021semantics}.
The speaker must describe its observed event to the listener, who is responsible for taking appropriate action in the environment according to the event so as to continue executing the associated task.
The action taken by the listener and the environment's dynamics will determine the next observed event. For semantic communication between the speaker and listener, we consider the existence of \emph{a common language} that is built using a \emph{set of beliefs} between the speaker the listener, known to both speaker and listener\footnote{The existence of such a language is a basic assumption commonly found in prior works \cite{seo2021semantics}.}.
{\color{black}For each observed event, there exists at least one representation in the belief set which can perfectly describe it. The perfect description of an event means that, by using the beliefs in such a description, the listener can successfully reconstruct the observed event. Precisely, a \emph{perfect description} of each event must have sufficient information (through the beliefs) about the event and, thus, it can be used for explaining the event to a remote listener thus allowing it to successfully reconstruct the event.}

To improve the communication resource (belief in our model) efficiency, the speaker and listener must cooperate to describe each event observed by the speaker. 
In other words, the speaker should transmit an abstract description of the observed event which is a subset of the perfect descriptor of the event, and the listener should infer what was sent by completing the received description, in a way to obtain a perfect descriptor of the observed event. The listener can complete the received description by adding new and appropriate beliefs from the belief set. We call these added beliefs the \emph{inferred description} of the listener. Therefore, the final, \emph{completed description} of each event will be, a combination of the speaker's transmitted description and the listener's inferred description. 
The listener utilizes the completed description to reconstruct the observed event.
Given its reconstructed event, the listener must take specific actions to steer the system in the right direction to complete the ongoing task. When the listener can successfully reconstruct the observed event, 
it can take a proper action in the environment regarding the observed event and immediately steer the ongoing task to its goal without delay.
However, if the reconstructed event does not match the observed event, the listener will take an incorrect action thus delaying the ongoing task. The speaker and the listener can perceive the correctness of the transmitted and inferred description for each event, according to the average task execution time.

\begin{figure*}[!t]
\centering
\begin{tikzpicture}[scale=0.65, every node/.style={scale=0.65}]

\node at (12,4)
    {\includegraphics[scale = 0.45]{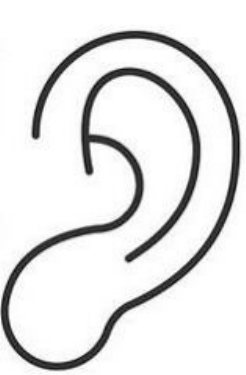}}; 
    
\node at (-0.6,4)
    {\includegraphics[scale = 0.32]{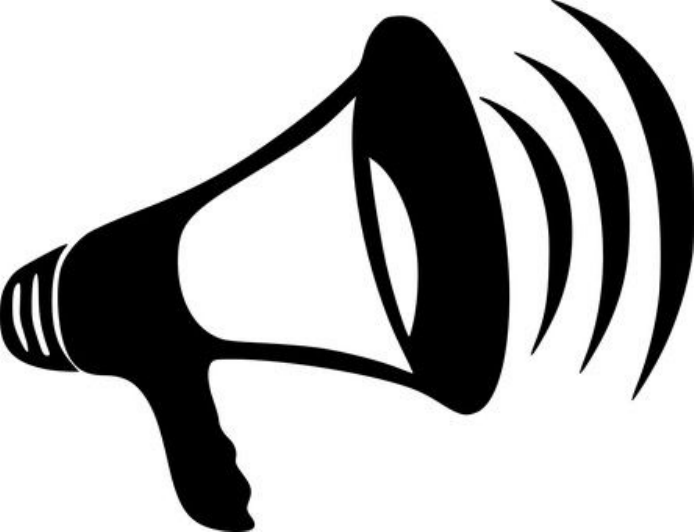}};
    
\node at (3,1.2)
    {\includegraphics[scale = 0.15]{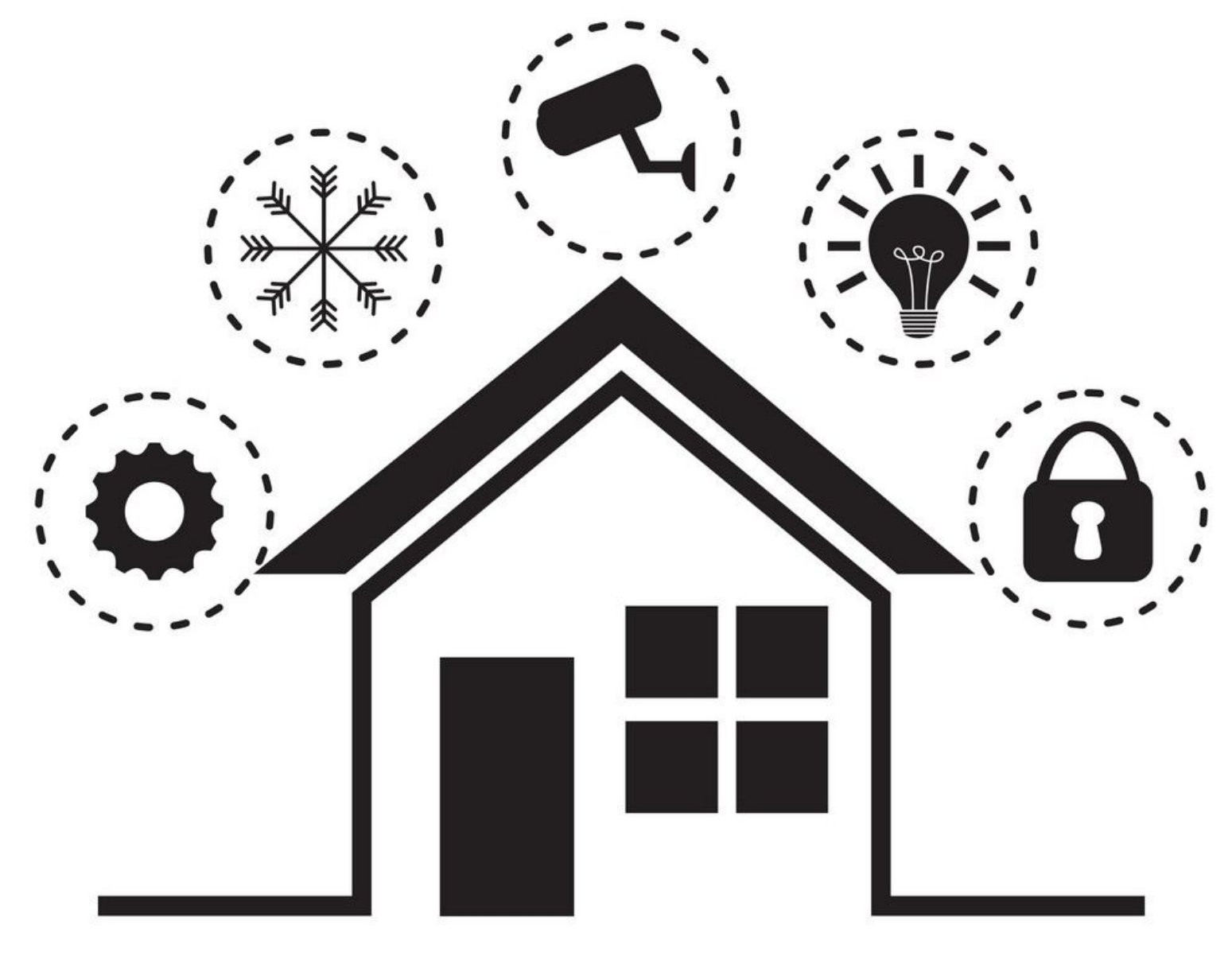}};
    
\node at (9,0.7)
    {\includegraphics[scale = 0.18]{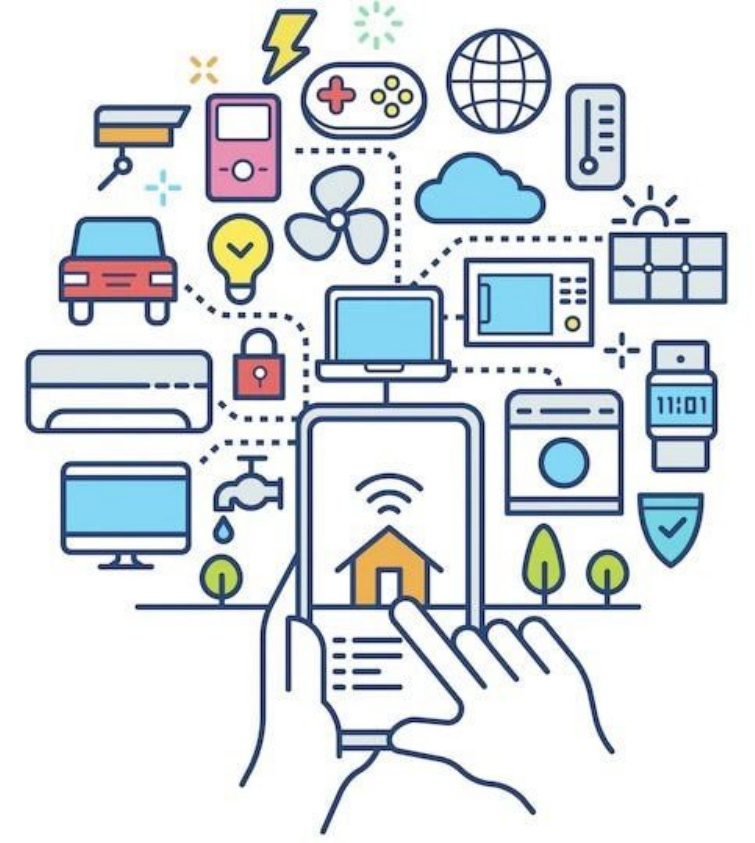}};
    
\node at (6.3,6)
    {\includegraphics[scale = 0.2]{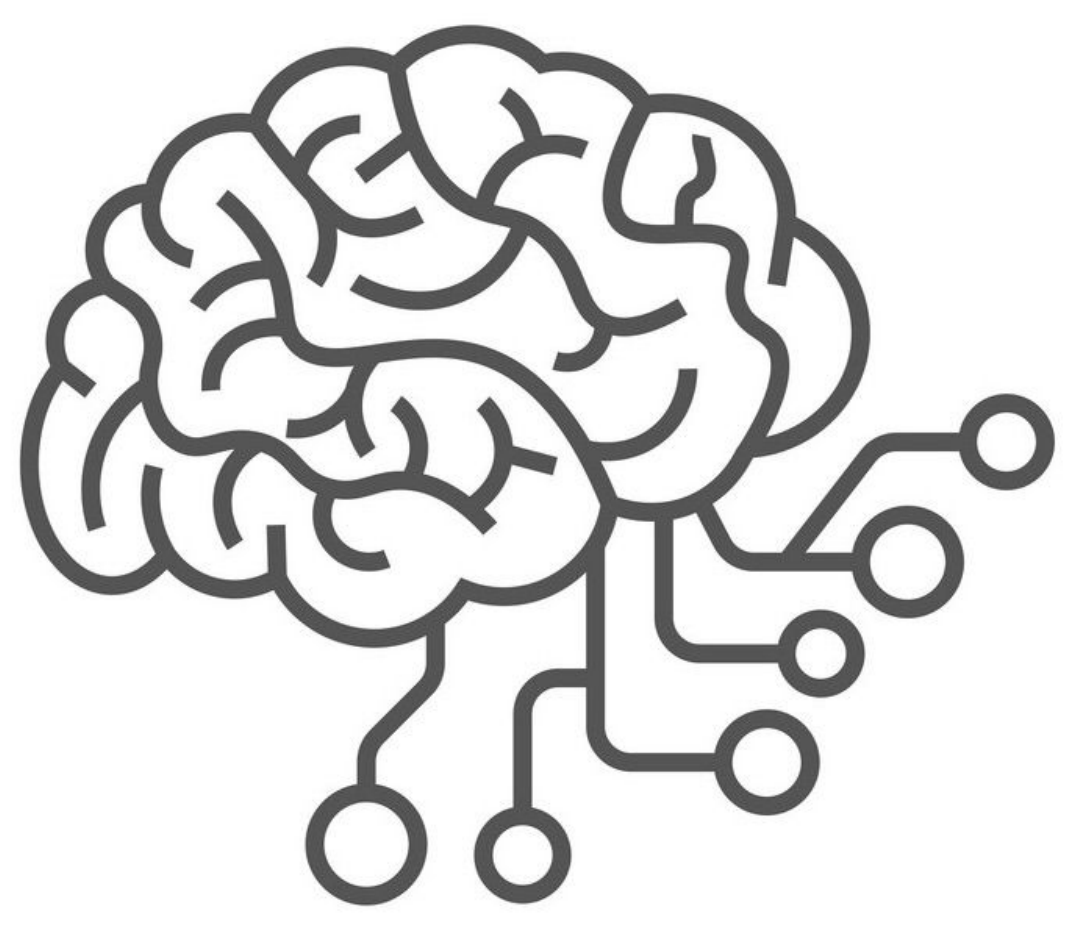}};
\draw [->, ultra thick] (4.4,6) -- (1.5,4.6);
\draw [->, ultra thick] (8.2,6) -- (11.2,4.6);  
\node[text width = 5cm] at (10,6.8) {\emph{Common language set}};

\draw[ultra thick, ->] (11.7, 2.1) arc (0:-90:1.6);
\draw[ultra thick, ->] (1, 0.6) arc (-90:-180:1.9);

\node[text width = 7.5cm] at (-0.7,1) {$e_{m,n+1}^S$ determined \\based on $e_{m,n}^S$ and $e_{m,n}^L$};
\node[text width = 5cm] at (4,3.6) {Speaker sends $\mathcal{B}_{m,n}^S$};
\draw [->, ultra thick] (1.4,4) -- (11.2,4);

\node[text width = 7cm] at (13,2.35) {Listener reconstructs $e_{m,n}^L$ using $\mathcal{B}_{m,n}$};
\node[text width = 7cm] at (9,3.25) {Listener infers $\mathcal{B}_{m,n}^L$ based on \\ $\mathcal{B}_{m,n}^S$ and $\mathcal{B}_{m,n} = \{\mathcal{B}_{m,n}^S, \mathcal{B}_{m,n}^L\}$};

\node[text width = 3cm] at (3.4, -0.6) {Environment};

\node[text width = 8cm] at (10, -0.6) {Listener takes action on the environment};
\draw [->, ultra thick] (8,0.5) -- (5,0.5);


\end{tikzpicture}
\vspace{-5.5mm}
\caption{\small {An example of a semantic goal-oriented communication between a speaker and a listener.}}
\label{Figs:System_Model}
\vspace{-10mm}
\end{figure*}
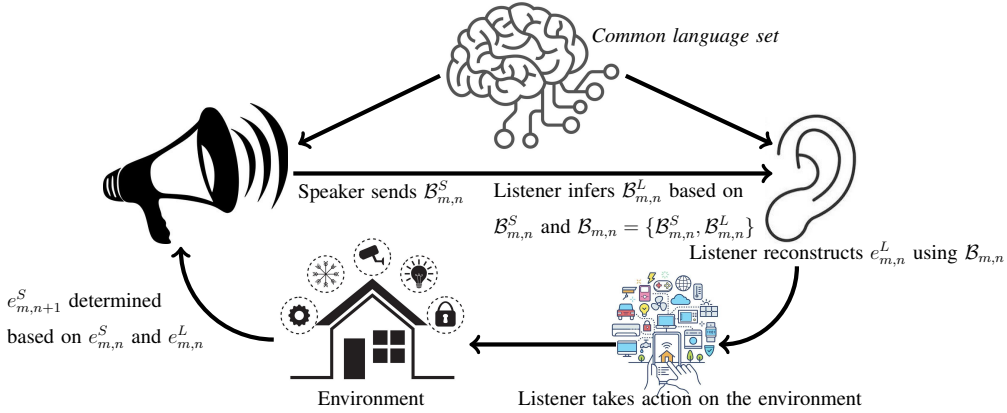

We divide time into different episodes, each of which is divided into equal time slots. At each episode $m$, a specific task $V_m$ is executed by the speaker-listener pair. At the beginning of slot $n$ in episode $m$, the speaker observes event $e_{m,n}^S$ as the state of the environment. We define $e_{m,n}^L$ as the reconstructed event of the listener, at time slot $n$ of episode $m$. Let $\mathcal{B}_{m,n}^S$, $\mathcal{B}_{m,n}^L$, and $\mathcal{B}_{m,n}$ be, respectively, the transmitted description of the speaker, the inferred description of the listener, and the completed description on the listener, at slot $n$ of episode $m$.
Fig. \ref{Figs:System_Model} illustrates the proposed goal-oriented semantic communication system. Here, the speaker observes an event \emph{$e_{m,n}^S$} and must describe it using $\mathcal{B}_{m,n}^S$. Then, the listener attempts to complete the received description based on its own inference and, then, reconstruct the observed event as \emph{$e_{m,n}^L$}. Finally, the listener takes an action in the environment based on the reconstructed event. 
It is worth noting that $e_{m,n}^L$ is considered to be equal to $e_{m,n}^S$, if and only if the listener's action based on $e_{m,n}^L$ is equal to the listener's action based on $e_{m,n}^S$.
Then, the environment transitions to the next event, which depends on the reconstructed event of the listener (which in fact determines its taken action) and the current observed event of the speaker.


\vspace{-7mm}
\subsection{Events and Tasks Definition}
Let $\mathcal{E} = \{\mathcal{E}_{\text{init}}, \mathcal{E}_{\text{mid}}, \mathcal{E}_{\text{fin}}\}$ be the set of all possible observed events
where $\mathcal{E}_{\text{init}}$, $\mathcal{E}_{\text{mid}}$, and $\mathcal{E}_{\text{fin}}$ are the set of initial, intermediary, and final events. Hence, the speaker starts each episode by observing an event from $\mathcal{E}_{\text{init}}$, and, then, it observes multiple events from $\mathcal{E}_{\text{mid}}$, and ends its monitoring of a given task by observing an event from $\mathcal{E}_{\text{fin}}$. Therefore, we can write $e_{m,1}^S \in \mathcal{E}_{\text{init}}$, $\{e_{m,2}^S, \ldots, e_{m,X_m-1}^S\} \in \mathcal{E}_{\text{mid}}$, and $e_{m,X_m}^S \in \mathcal{E}_{\text{fin}}$, where $X_m$ is the length of episode $m$ quantified in terms of number of time slots, and it is equal to \emph{the execution time of the task} performed during episode $m$.
Now, we can define a task based on the different events of the event set. Let $\mathcal{T}$ be the set of all $\nu$ tasks. Then, we can define  $\mathcal{O}^T$ as the TEC of task $T$, as follows:
\vspace{-3mm}
\begin{align}
\mathcal{O}^T = \Big\{\big(e^T_1, e^T_2, \ldots, e^T_{{L_T^{\text{max}}-1}}, e^T_{{L_T^{\text{max}}}} \big) \Big |
e^T_1 \in \mathcal{E}_{\text{init}}, \big( e^T_2, \ldots, e^T_{L_T^{\text{max}}-1}\big) \in  \mathcal{E}_{\text{mid}}, \, e^T_{L_T^{\text{max}}} \in \mathcal{E}_{\text{fin}} \Big\}, \label{Equation:Task_Def}
\end{align}
where $e^T_j$ is the $j^{th}$ event in the TEC of task $T$ and 
$L_T^{\text{max}}$ is the maximum length of task $T$ in number of time slots, which is fixed for each task but can vary from one task to another. Due to the randomness of the environment's dynamics, the length of task $T$ will vary between different episodes. 
{\color{black}The environment's dynamic includes all unknown factors in the environment for the speaker and listener that affect how the observed event of the speaker evolves from the current time slot to the next time slot (in addition to the action taken by the listener). These unknown factors of the environment are out of control of the speaker and listener and modeled by the transition model of the environment (matrices $\boldsymbol{P}$ and $\boldsymbol{\Tilde{P}}$).}
Let $L_T$ be a discrete random variable that captures the length of task $T$ across episodes. $L_T$ is distributed according to a probability mass function (PMF) $f_T(L) = \text{Prob}(L_T = L)$, where $L \in \big\{3, 4, \ldots, L_T^{\text{max}}\big\}$. The minimum value for $L$ is 3 because we assume that each task has at least one event from each of $\mathcal{E}_{\text{init}}$, $\mathcal{E}_{\text{mid}}$, and $\mathcal{E}_{\text{fin}}$. Also, we assume that the initial event of each task is fixed to differentiate among different tasks.
{\color{black}One simple example of our proposed model is the remote control of high-precision manufacturing in automated factories in which there are multiple devices (e.g., cameras and robots) distributed in the factory to observe the state of the production line. The devices should transmit their captured data (equivalent to the event in our model) to a coordinator (e.g., an aggregator or a base station). The coordinator is responsible for taking control action in the factory based on the received information. Examples of tasks in this application include the coordination of a group of robots and devices to build a product, real-time monitoring of the production line, and resolution of a deficiency in the production line.}

Next, we explain the relation between $L_T$ and the length of episode $m$, $X_m$, which is the execution time of the performed task in episode $m$. 
Assume that the executed task in episode $m$ is task $T$. Due to the probabilistic nature of $L_T$, when the listener takes a proper action for each observed event during episode $m$, then we expect to have $X_m = \mathbb{E}[L_T]$, where $\mathbb{E}[L_T]$ is the expectation of $L_T$. For taking a proper action, the listener must successfully reconstruct each observed event during episode $m$.
However, the speaker and listener do not know the perfect descriptions of the events, and they need to learn them. During learning, the speaker's transmitted descriptions can be imperfect or incomplete, and the listener cannot perfectly infer and complete the received description, and thus, the completed description may not perfectly describe the observed events. In this case, the listener cannot successfully reconstruct the observed events, the action taken by the listener will not be accurate, and the ongoing task will be delayed. 
When executing task $T$ in episode $m$, we have $e_{m,1}^S = e^T_1$. Assume that at the first slot of episode $m$, we have $e_{m,1}^L = e_{m,1}^S$, then $e_{m,2}^S \in \{e^T_2, \ldots, e^T_{L_T^{\text{max}}-1}\}$. However, if the listener cannot successfully reconstruct the observed event in the second slot, and $e_{m,2}^L \neq e_{m,2}^S$, then, for the next observed event, we have $e_{m,3}^S \in \mathcal{E}$ and $e_{m,3}^S \notin \{e^T_3, \ldots, e^T_{L_T^{\text{max}}}\}$, which means that the task is delayed. Hence, in this case, during learning, we anticipate that the execution time of the executed task in episode $m$, will be greater than its expected length and, thus, $X_m > \mathbb{E}[L_T]$.
\vspace{-7mm}
\subsection{Common Language Belief Set and Semantic Communication Costs}
\vspace{-3mm}
We consider a basic \emph{common language set} between the speaker and listener for describing the observed events of the speaker. Let $\mathcal{B}$ be the set of $B$ beliefs of this language. Each belief $b \in \mathcal{B}$ is a unique feature that can be used for event description. Thus, the belief set can be different for different applications. Intuitively, the belief set for a specific application can be determined by performing a semantic extraction procedure that is beyond the scope of this work. We assume the speaker and listener have same perception about the mapping between beliefs and the events and $\mathcal{B}$ to be a fixed input variable \cite{farshbafan2021common}. 
One possibility for the beliefs is to represents words in a text, which is efficient for the description of multimedia data (e.g., image captioning) and it can significantly reduce the usage of communication resources between the speaker and listener. In an image description task, we can consider the transmission of image objects and their specific characteristics using text data. In this example, our beliefs can be the set of objects and their characteristics. For applications in which the text data type is not sufficient for describing the events, we can consider the existence of the beliefs with other data types. Therefore, each element of the belief set can capture a certain data type like text, image, and sensory information. The combination of beliefs with different data types can be used for event descriptions.
Note that the beliefs can then be transmitted using classical wireless techniques.
{\color{black}It is worth noting that in contrast to most DL-driven semantic (goal-oriented) communication systems \cite{zhang2022unified, xie2021task, xie2022task}, which are based on extracting the features of the data (e. g., image or audio) using a neural network, in our proposed method, the beliefs of the belief set, which are used for event description, are semantic meanings related to the events. Also, they should have considered a presence of a dynamic environment, as considered in our model, which can affect the task execution procedure. Finally, our proposed model jointly optimizes the task execution and wireless resource metrics, which is not the case in the existing DL-driven works.}

{\color{black}The amount of information that the different elements of the belief set can convey is different. Precisely, some elements of the belief set have general information about the environment and can be used for the perfect description of a large number of events. On the other hand, some other elements of the belief set describe more specific features about the environment and, thus, they can be used for the perfect description of a small number of events compared to the first group of beliefs. The first and the second groups of beliefs are called superordinate and subordinate concepts, respectively \cite{higgins2017scan}. Also, in practice, some features are inconsistent and, thus, they cannot be simultaneously used for event description. For example, in an image description task, we can consider an animal in the image as general information (high-level belief) compared to the exact species of the animal, which provides more specific information (low-level belief). Also, the specific behavior of that species (e.g., jumping, running or sitting) conveys more information compared to the exact species. This is just a simple example and our model can be used for more complex ones. To model all of these constraints of the belief set including the different data types, their relations, and the amount of semantic information that they can convey, we propose a \emph{hierarchical structure} for the belief set \cite{higgins2017scan}}. 
In this structure, the beliefs can be categorized in \emph{different levels, based on their amount of semantic information, and their data type}\footnote{This hierarchical structure can also be seen as a way to reflect the causality relations between different beliefs.}.
{\color{black} Here, we considered simple relations to combine the beliefs for event description. To model complicated relations between the beliefs to combine them for event description, we can use techniques such as factor graphs \cite{loeliger2004introduction}.}
In a hierarchical structure, the beliefs in the first level convey general information about the events and, thus, the amount of semantic information conveyed by the beliefs in the first level is less than that of the beliefs in other levels.
By going deeper in the hierarchical structure, the amount of semantic information conveyed by the beliefs increases as deeper beliefs contain more detailed information about the events. According to this structure, the belief set can be defined as follows:
\vspace{-3mm}
\begin{align}
    \mathcal{B} = \Big \{\mathcal{B}_1, \mathcal{B}_2, \ldots, \mathcal{B}_K \Big\}, \quad \mathcal{B}_k = \Big\{b_{k,1}, b_{k,2}, \ldots, b_{k, B_k} \Big\}, \quad \mathcal{B}_k \cap \mathcal{B}_{k^{\prime}} = \emptyset, \; \forall \, 1 \leq k, k^{\prime} \leq K,
    \label{Equation:Belief_Def}
\end{align}
where $\mathcal{B}_k$ is the belief set of level $k$ of the hierarchy, $B_k$ is the number of beliefs in level $k$, and $K$ is the total number of levels of the belief set hierarchy. The total number of beliefs will be $B = \sum_{k=1}^{K}{B_k}$. The intersection of the belief sets of two different levels will be empty since each belief can only belong one level of the hierarchy. 
Let $C_{k,u}^S$ be the cost incurred by the speaker for transmitting belief $u$ of level $k$ and $C_{k,u}^L$ be the cost incurred by the listener for inferring and adding belief $u$ of level $k$ to a received description. The inference cost can be viewed as the computational cost that the listener requires to add beliefs and compute the completed description.
The cost of each belief is determined based on the amount of the semantic information that it can convey. As such, we consider $C_{k,u}^S$ and $C_{k,u}^L$ to be increasing functions of semantic information. 
The increasing costs with the hierarchy are further justified by the fact the more detailed beliefs require more bits for transmission in a classic communication system.

According to the definition of the belief set, which includes different features of the events, there could be multiple subsets of the belief set that can perfectly describe a given event.
Since the belief set has a hierarchical structure, then, the perfect descriptor of each event is also hierarchical. Precisely, a perfect description of an event only contains one belief from each level of the hierarchy. 
Let $\mathcal{B}_{e}^P$ be the set of the perfect descriptors of event $e \in \mathcal{E}$, defined as follows:
\vspace{-9mm}
\begin{align}
  \mathcal{B}_{e}^P &= \Big\{\mathcal{B}_h \subset \mathcal{B} \Big| \text{Listener can successfully reconstruct } e \text{ based on } \mathcal{B}_h\Big\},  \label{Equation:Perfect_Description_Def} \\
  \mathcal{B}_h &= \Big\{b_{1,u_1}, b_{2,u_2}, \ldots, b_{K_h,u_{K_h}} \Big\},\quad b_{k, u_k} \in \mathcal{B}_{k}, \quad 2 \leq K_h \leq K,
  \notag
\end{align}
where $\mathcal{B}_h$ is a perfect, hierarchical descriptor of event $e$, $K_h$ is the number of the hierarchy levels, which is different for each event, and $b_{k, u_k}$ is a belief in the level $k$ of the hierarchy.

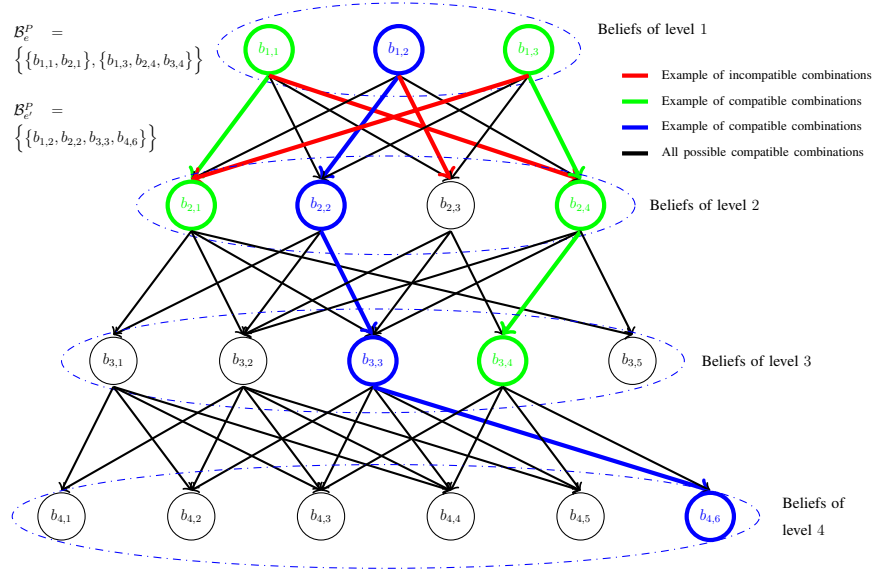
\begin{figure*}[!t]
\centering
\begin{tikzpicture}[scale=0.69, every node/.style={scale=0.49}]

\node[draw, align = center, text width=0.8cm, color = green, circle, ultra thick] at (6.5,11) {$b_{1,1}$};
\node[draw, align = center, text width=0.8cm, color = blue, circle, ultra thick] at (9,11) {$b_{1,2}$};
\node[draw, align = center, text width=0.8cm, color = green, circle, ultra thick] at (11.5,11) {$b_{1,3}$};

\draw[blue, dash dot, thin] (9, 11) ellipse (3.5 and 0.9);
\node[text width=3cm] at (13.9,11.4) 
    {Beliefs of level $1$};

\draw [->, color=green, ultra thick] (6.5,10.5) -- (5,8.5);
\draw [->, color=black, thick] (6.5,10.5) -- (7.5,8.5);
\draw [->, color=black, thick] (6.5,10.5) -- (10,8.5);
\draw [->, color=red, ultra thick]   (6.5,10.5) -- (12.5,8.5);

\draw [->, color=black, thick] (9,10.5) -- (5,8.5);
\draw [->, color=blue, ultra thick] (9,10.5) -- (7.5,8.5);
\draw [->, color=black, thick] (9,10.5) -- (12.5,8.5);
\draw [->, color=red, ultra thick]   (9,10.5) -- (10,8.5);

\draw [->, color=black, thick] (11.5,10.5) -- (7.5,8.5);
\draw [->, color=black, thick] (11.5,10.5) -- (10,8.5);
\draw [->, color=green, ultra thick] (11.5,10.5) -- (12.5,8.5);
\draw [->, color=red, ultra thick]   (11.5,10.5) -- (5,8.5);

\node[draw, align = center, text width=0.8cm, color = green, circle, ultra thick] at (5,8) {$b_{2,1}$};
\node[draw, align = center, text width=0.8cm, color = blue, circle, ultra thick] at (7.5,8) {$b_{2,2}$};
\node[draw, align = center, text width=0.8cm, color = black, circle] at (10,8) {$b_{2,3}$};
\node[draw, align = center, text width=0.8cm, color = green, circle, ultra thick] at (12.5,8) {$b_{2,4}$};

\draw[blue, dash dot, thin] (8.75, 8) ellipse (4.8 and 0.95);
\node[text width=3cm] at (14.9,8) 
    {Beliefs of level $2$};
    

\draw [->, color=black, thick] (5,7.5) -- (3.5,5.5);
\draw [->, color=black, thick] (5,7.5) -- (6,5.5);
\draw [->, color=black, thick] (5,7.5) -- (8.5,5.5);
\draw [->, color=black, thick] (5,7.5) -- (13.5,5.5);

\draw [->, color=black, thick] (7.5,7.5) -- (3.5,5.5);
\draw [->, color=black, thick] (7.5,7.5) -- (6,5.5);
\draw [->, color=blue, ultra thick] (7.5,7.5) -- (8.5,5.5);

\draw [->, color=black, thick] (10,7.5) -- (6,5.5);
\draw [->, color=black, thick] (10,7.5) -- (8.5,5.5);
\draw [->, color=black, thick] (10,7.5) -- (11,5.5);

\draw [->, color=black, thick] (12.5,7.5) -- (6,5.5);
\draw [->, color=black, thick] (12.5,7.5) -- (8.5,5.5);
\draw [->, color=green, ultra thick] (12.5,7.5) -- (11,5.5);
\draw [->, color=black, thick] (12.5,7.5) -- (13.5,5.5);

\node[draw, align = center, text width=0.8cm, color = black, circle] at (3.5,5) {$b_{3,1}$};
\node[draw, align = center, text width=0.8cm, color = black, circle] at (6,5) {$b_{3,2}$};
\node[draw, align = center, text width=0.8cm, color = blue, circle, ultra thick] at (8.5,5) {$b_{3,3}$};
\node[draw, align = center, text width=0.8cm, color = green, circle, ultra thick] at (11,5) {$b_{3,4}$};
\node[draw, align = center, text width=0.8cm, color = black, circle] at (13.5,5) {$b_{3,5}$};

\draw[blue, dash dot, thin] (8.5, 5) ellipse (6 and 1);
\node[text width=3cm] at (15.9,5) 
    {Beliefs of level $3$};

\draw [->, color=black, thick] (3.5,4.5) -- (2.5,2.5);
\draw [->, color=black, thick] (3.5,4.5) -- (5,2.5);
\draw [->, color=black, thick] (3.5,4.5) -- (7.5,2.5);
\draw [->, color=black, thick] (3.5,4.5) -- (10,2.5);

\draw [->, color=black, thick] (6,4.5) -- (2.5,2.5);
\draw [->, color=black, thick] (6,4.5) -- (5,2.5);
\draw [->, color=black, thick] (6,4.5) -- (7.5,2.5);
\draw [->, color=black, thick] (6,4.5) -- (10,2.5);
\draw [->, color=black, thick] (6,4.5) -- (12.5,2.5);

\draw [->, color=black, thick] (8.5,4.5) -- (5,2.5);
\draw [->, color=black, thick] (8.5,4.5) -- (7.5,2.5);
\draw [->, color=black, thick] (8.5,4.5) -- (10,2.5);
\draw [->, color=black, thick] (8.5,4.5) -- (12.5,2.5);
\draw [->, color=blue, ultra thick] (8.5,4.5) -- (15,2.5);

\draw [->, color=black, thick] (11,4.5) -- (7.5,2.5);
\draw [->, color=black, thick] (11,4.5) -- (10,2.5);
\draw [->, color=black, thick] (11,4.5) -- (12.5,2.5);
\draw [->, color=black, thick] (11,4.5) -- (15,2.5);

\node[draw, align = center, text width=0.8cm, color = black, circle] at (2.5,2) {$b_{4,1}$};
\node[draw, align = center, text width=0.8cm, color = black, circle] at (5,2) {$b_{4,2}$};
\node[draw, align = center, text width=0.8cm, color = black, circle] at (7.5,2) {$b_{4,3}$};
\node[draw, align = center, text width=0.8cm, color = black, circle] at (10,2) {$b_{4,4}$};
\node[draw, align = center, text width=0.8cm, color = black, circle] at (12.5,2) {$b_{4,5}$};
\node[draw, align = center, text width=0.8cm, color = blue, circle, ultra thick] at (15,2) {$b_{4,6}$};

\draw[blue, dash dot, thin] (8.75, 2) ellipse (7.2 and 1);
\node[text width=2cm] at (17.1,2) 
    {Beliefs of level $4$};

\node[text width = 4cm] at (3,11) {$\mathcal{B}_{e}^P = \Big\{ \big\{b_{1,1}, b_{2,1} \big\},  \big\{b_{1,3}, b_{2,4}, b_{3,4}\big\} \Big\}$};

\node[text width = 4cm] at (3,9.5) {$\mathcal{B}_{e^{\prime}}^P = \Big\{\big\{b_{1,2}, b_{2,2}, b_{3,3}, b_{4,6}\big\}\Big\}$};

\draw [-, color=red, ultra thick] (13.3,10.5) -- (13.8,10.5);
\node[text width=6cm] at (16.2,10.5) 
    {\small{Example of incompatible combinations}};

\draw [-, color=green, ultra thick] (13.3,10) -- (13.8,10);
\node[text width=6cm] at (16.2,10) 
    {\small{Example of compatible combinations}};
    
\draw [-, color=blue, ultra thick] (13.3,9.5) -- (13.8,9.5);
\node[text width=6cm] at (16.2,9.5) 
    {\small{Example of compatible combinations}};
    
\draw [-, color=black, ultra thick] (13.3,9) -- (13.8,9);
\node[text width=6cm] at (16.2,9) 
    {\small{All possible compatible combinations}};

\end{tikzpicture}
\vspace{-5mm}
\caption{\small{An example of the hierarchical structure of the belief set including four levels.}}
\label{Fig:Hierarchical_Example}
\vspace{-12mm}
\end{figure*}


Fig. \ref{Fig:Hierarchical_Example} shows an example of the hierarchical belief set, $\mathcal{B}$, that includes four levels, where there are $3$, $4$, $5$, and $6$ beliefs in the first, second, third, and fourth levels, respectively. The red edges represent \emph{incompatible combinations} (i.e., not useful for event description) between the beliefs of the first and second levels. Meanwhile, the black, green, and blue edges represent several possible \emph{compatible combinations} between the beliefs of different levels. Also, Fig. \ref{Fig:Hierarchical_Example} illustrates the perfect description of two events $e \in \mathcal{E}$ and $e^{\prime} \in \mathcal{E}$. Event $e$ has two perfect descriptors, as captured by the green edges. The first and second perfect descriptors of $e$ include two and three beliefs, respectively. Event $e^{\prime}$ has one perfect descriptor, which is shown using blue edges and includes four beliefs. 
Here, in contrast to \cite{farshbafan2021common}, adding more beliefs to a perfect description of an event $e$ can make that description imperfect for describing $e$.
For example, in Fig. \ref{Fig:Hierarchical_Example}, $\big\{b_{1,1}, b_{2,1}\big\}$ is considered as a perfect descriptor of event $e$. In contrast, for this same figure, description $\big\{b_{1,1}, b_{2,1}, b_{k,u}\big\}$ for any values of $k$ and $u$ is not necessarily a perfect descriptor of $e$.

We next explain the cost of each event reconstruction by the speaker and listener. Here, we have a dual-objective optimization problem, including minimizing the transmission cost of the speaker and the inference cost of the listener. This dual-objective optimization can be modeled in a variety of ways. One popular approach is to use scalarization \cite{gunantara2018review} which we adopt here. The scalarization method incorporates multi-objective functions into a scalar fitness function \cite{murata1996multi}, and it is known as a meaningful way to represent such problems.
Let $C_{m,n}^T$ be the total cost of the completed description used at slot $n$ of episode $m$, which can be computed as follows:
\vspace{-4.5mm}
\begin{alignat}{3}
    C_{m,n}^T &= \alpha \cdot C_{m,n}^S + (1-\alpha) \cdot C_{m,n}^L, \label{Equation:Slot_Cost} \\ 
    C_{m,n}^S &= \textstyle\sum_{k=1}^{K}{\sum_{u=1}^{B_k}{C_{k,u}^S \cdot x_{m,n,k,u}^S}}, \quad \;\,
    C_{m,n}^L &&= \textstyle\sum_{k=1}^{K}{\sum_{u=1}^{B_k}{C_{k,u}^L \cdot x_{m,n,k,u}^L}},  \label{Equation:Slot_Cost_Speaker_Listener}\\
    x_{m,n,k,u}^S &= \begin{cases} 1, & b_{k,u} \in \mathcal{B}^S_{m,n},   \\ 0, & b_{k,u} \in \mathcal{B}^S_{m,n}, \end{cases} \quad \quad
    x_{m,n,k,u}^L &&= \begin{cases} 1, & b_{k,u} \in \mathcal{B}^L_{m,n},   \\ 0, & b_{k,u} \in \mathcal{B}^L_{m,n}. \end{cases}
\end{alignat}

Here, $x_{m,n,k,u}^S$ is a binary decision variable that indicates whether belief $u$ of level $k$ of the hierarchy is used in $\mathcal{B}_{m,n}^S$.
Similarly, $x_{m,n,k,u}^L$ is a binary decision variable that indicates whether belief $u$ of level $k$ of the hierarchy is used in $\mathcal{B}_{m,n}^L$.
$C_{m,n}^S$ is the incurred cost of the speaker for transmitting $\mathcal{B}_{m,n}^S$ and $C^L_{m,n}$ is the incurred cost of the listener for inferring $\mathcal{B}_{m,n}^L$ at slot $n$ of episode $m$.
$\alpha$ is a design parameter for making a compromise between the costs of the speaker and the listener. 
Here, we assume that the listener can only complete the speaker's transmitted description by adding new beliefs based on its own inference. In other words, the listener does not ignore any beliefs transmitted by the speaker, and it uses all of them for event reconstruction.
\vspace{-14mm}
\subsection{Environment State Evolution}
\vspace{-2mm}
The environment's state at each time slot, captured by the observed event of the speaker $e_{m,n}^S$, is dependent on the environment's state and the action taken by the listener in the environment, at the previous slot.
The speaker describes its observation by transmitting $\mathcal{B}_{m,n}^S$ to the listener. 
Then, the listener infers and completes $\mathcal{B}_{m,n}^S$ by adding new beliefs, captured by $\mathcal{B}_{m,n}^L$. Finally, the listener reconstructs the observed event of the speaker, based on the completed description. Thus, the reconstructed event $e_{m,n}^L$ by the listener at each time slot, can be computed as follows:
\vspace{-8mm}
\begin{align}
    e_{m,n}^L = \begin{cases} e_{m,n}^S, & \mathcal{B}_{m,n} \in \mathcal{B}^P_{e},  \\ e \in \big\{\mathcal{E} \setminus e_{m,n}^S\big\}, & \mathcal{B}_{m,n} \notin \mathcal{B}^P_{e}, \end{cases} 
\end{align}
where $\mathcal{B}_{m,n} = \{\mathcal{B}_{m,n}^S, \mathcal{B}_{m,n}^L\}$ is the completed description and $\{\mathcal{E} \setminus e_{m,n}^S\}$ is the set of all events except $e_{m,n}^S$. To capture the environment evolution from one slot to another, we must define the transition probabilities between different events as functions of the reconstructed event by the listener, $e_{m,n}^L$. Let $p_{j_1, j_2}$ be the transition probability from $e_{m,n}^S = e_{j_1}$ to $e_{m,n+1}^S = e_{j_2}$, given by:
\vspace{-2mm}
\begin{align}
    p_{j_1, j_2} = \text{Pr}\Big(e_{m,n}^S=e_{j_1} \rightarrow e_{m,n+1}^S = e_{j_2} \Big| e_{m,n}^L\Big) = \begin{cases} \boldsymbol{P}(j_1,j_2) & e_{m,n}^L = e_{j_1}  \\ \boldsymbol{\Tilde{P}}(j_1,j_2) & e_{m,n}^L \neq e_{j_1}, \end{cases}
    \label{Equation:Transition_Prob}
\end{align}
where $\boldsymbol{P}$ is the transition probability matrix of the environment's events when the listener can take proper action regarding the observed event. We expect $\boldsymbol{P}$ to be a sparse matrix because, when the listener can take proper action regarding the state of the environment, the next state is one of the events of the ongoing task's TEC.
$\boldsymbol{\Tilde{P}}$ is the transition probability matrix when the speaker does not take the right action regarding the observed event because the listener cannot successfully reconstruct the observed event. Hence, we expect $\boldsymbol{\Tilde{P}}$ to be a random matrix that has more nonzero elements compared to $\boldsymbol{P}$. This is due to the fact that, when the listener cannot take a proper action for a given state of the environment, the next state will have higher randomness.
Precisely, we assume that, when the listener cannot take proper action, the probability distribution of the next observed event is completely random. In such scenarios, the transition probability can be uniform over all possible events. In this case, there is no difference between the taken actions in terms of the transition probability matrix in the second case of \eqref{Equation:Transition_Prob} and all of the improper actions that lead to a random behavior in terms of the transition probability matrix.
\vspace{-8mm}
\subsection{Task Execution Metrics}
\vspace{-2mm}
We consider three key performance metrics for our semantic communication system:
\subsubsection{Task Execution Cost}
We know that there can be multiple subsets of the belief set which can perfectly describe each event $e$, as captured by $\mathcal{B}_{e}^P$ and defined in \eqref{Equation:Perfect_Description_Def}. To differentiate between the perfect descriptors of an event, in  \eqref{Equation:Slot_Cost}, we defined the cost of the used descriptor at each time slot. We now compute the cost $C_m$ of the executed task in episode $m$. 
This cost will be a function of the incurred cost of the speaker and listener during episode $m$, as follows:
\vspace{-4.5mm}
\begin{align}
     C_m &= \alpha \cdot C_{m}^S + (1- \alpha) \cdot C_{m}^L = \alpha \cdot  \textstyle\sum_{n=1}^{X_m}{C_{m,n}^S} + (1- \alpha) \cdot \ \textstyle\sum_{n=1}^{X_m}{C_{m,n}^L} \label{Equation:Episode_Cost} \\
      &= \alpha \cdot \textstyle\sum_{n=1}^{X_m}{  \sum_{k=1}^{K}{\sum_{u=1}^{B_k}{C_{k,u}^S \cdot x_{m,n,k,u}^S}}} + (1- \alpha) \cdot \textstyle\sum_{n=1}^{X_m}{\sum_{k=1}^{K}{\sum_{u=1}^{B_k}{C_{k,u}^L \cdot x_{m,n,k,u}^L}}} \notag.
\end{align}

{\color{black}In our model, the listener is a remote node and it is not the observer of the environment. For example, the listener can be a base station or an aggregator. Therefore, we can assume that the listener is computationally capable, and it possesses a wide range of inference capabilities.}
\subsubsection{Task Execution Time}
We know that the length of each task $T$ is a random variable. Thus, for each task $T$, we should consider the \emph{mean} task execution time over different episodes, $\Gamma_T$. We consider the length of each episode, $X_m$, as the task execution time for the task executed during episode $m$. To compute $\Gamma_T$, we should calculate the average of the episode length $\Gamma_T = \mathbb{E}\big[X_m\big|V_m =T\big]$, during which task $T$ is executed,
where $V_m$ is the index of the task executed in episode $m$ and the expectation is over different episodes. The speaker and listener must learn the perfect descriptors of each event and, during the learning process, they cannot perfectly describe the observed event hence delaying the tasks. Thus, during learning, we have $\Gamma_T > \mathbb{E}[L_T]$, where $\mathbb{E}[L_T]$ is the expected length of task $T$ calculated over $f_T$.
However, as the learning evolves over time and the speaker and listener gradually learn the perfect description of the events, we expect that $\lim_{M\rightarrow \infty} [\Gamma_T] \rightarrow \mathbb{E}[L_T]$, where $M$ is the number of episodes.

\subsubsection{Belief Efficiency}
In semantic communications, the main resource of the speaker is the beliefs, which can be considered as the equivalent of bandwidth in classical communication. Spectral efficiency is the main metric to optimize bandwidth usage in classical communication. We now optimize the belief usage by defining a similar metric. Let $U_b$ be the belief efficiency. 
Let $U_{b,T}$ and $U_{b,m}$ be the belief efficiency for task $T$ and belief efficiency of episode $m$, respectively.
To compute $U_{b,m}$, we should consider the total number (across all time slots) of beliefs that the speaker uses to describe the observed events during episode $m$, and inverse it, as follows:
\vspace{-3.5mm}
\begin{align}
     U_{b,m}=\dfrac{1}{\sum_{n=1}^{X_m}{W_{m,n}^S}}, \quad  W_{m,n}^S = \textstyle\sum_{k=1}^{K}{\sum_{u=1}^{B_k}{x_{m,n,k,u}^S}}, \label{Equation:Episode_Belief_Efficiency}
\end{align}
where $W_{m,n}^S$ is the total number of beliefs transmitted for describing the observed event, at slot $n$ of episode $m$.
For computing the belief efficiency, we only consider the beliefs sent by the speaker while the inferred beliefs of the listener are ignored because they are not transmitted over a wireless channel. To compute $U_{b,T}$, we should calculate the expectation of $U_{b,m}$ in the episodes during which task $T$ is executed, i.e., $U_{b,T}= \mathbb{E}\big[U_{b,m}\big|V_m =T\big]$. This is because the required number of the beliefs for executing a task can randomly vary across episodes since the number of events in each task's TEC is a random variable. Finally, the belief efficiency $U_b = \mathbb{E} \big[U_{b,T}\big]$ will be the expectation of $U_{b,T}$ over different tasks.
\vspace{-6mm}
\section{Problem Formulation and Proposed CL Algorithm}
\vspace{-2mm}
\label{Section:Proposed_Method}
\subsection{Problem Formulation}
\vspace{-2mm}
\label{Subsec_Optimization}
Given the metrics defined in Section II, we can now formally
pose the problem of determining the perfect and abstract description of the events, as follows:
\vspace{-2mm}
\begin{subequations}
\begin{align} 
\hspace{-2mm}\min_{\boldsymbol{X}_{m,n}^S, \boldsymbol{X}_{m,n}^L \in {\mathcal{A}}} & \;\Big [\lim_{M \to \infty} \textstyle\sum_{m=1}^{M} \gamma^m C_m  \Big ], \label{Equation:Objective}\\
 \text{s. t.} &\; \lim_{M\rightarrow \infty} \big[\Gamma_T\big] = E\big\{L_T\big\}, \quad T = 1,2,\ldots, \nu, 
 \label{Equation:Constraint_Time} \\
 & \; \lim_{M\rightarrow \infty} \big[U_b\big] \geq U_b^{\text{min}},
 \label{Equation:Constraint_Belief} \\
 & \; 
 \boldsymbol{X}_{m,n}^S[k,u] + \boldsymbol{X}_{m,n}^L[k,u] \leq 1, \quad u = 1,2,\ldots, B_k, \; k = 1,2,\ldots,K,
 \label{Equation:Constraint_3} \\
 & \; 
\textstyle \sum_{u=1}^{B_k}{\boldsymbol{X}_{m,n}^S[k,u] + \boldsymbol{X}_{m,n}^L[k,u]} \leq 1, \quad  k = 1,2,\ldots,K,
 \label{Equation:Constraint_4} \\
 & \; \textstyle\sum_{u=1}^{B_k}{\boldsymbol{X}_{m,n}^S[k+1,u] + \boldsymbol{X}_{m,n}^L[k+1,u]} \leq \textstyle\sum_{u=1}^{B_k}{\boldsymbol{X}_{m,n}^S[k,u] + \boldsymbol{X}_{m,n}^L[k,u]}, 
\label{Equation:Constraint_5}
\end{align}
\label{Equation:Optimization}
\end{subequations}
\hspace{-2mm}where the objective function is the discounted sum of the total costs of both the speaker and listener during an infinite horizon and $\gamma$ is a discount factor. $\boldsymbol{X}_{m,n}^S = \big[x_{m,n,k,u}^{S}\big]_{K \times B_K}$ and $\boldsymbol{X}_{m,n}^L = \big[x_{m,n,k,u}^{L}\big]_{K \times B_K}$ are, respectively, the matrix format decision variables of the speaker and listener at slot $n$ of episode $m$, over different values $k$ and $u$. ${\mathcal{A}}$ is the set of all possible binary matrices of size $K \times B_K$. $C_m$ is a function of $\boldsymbol{X}_{m,n}^S$ and  $\boldsymbol{X}_{m,n}^L$ according to \eqref{Equation:Episode_Cost}, but, for notational simplicity, we do not show this dependence in \eqref{Equation:Objective}.
\eqref{Equation:Constraint_Time} is used to minimize the task execution time. More precisely, this constraint motivates the speaker and listener to find the perfect description of each event. \eqref{Equation:Constraint_Belief} is used to optimize the belief efficiency. Note that \eqref{Equation:Constraint_Time} and \eqref{Equation:Constraint_Belief} depend on $\boldsymbol{X}_{m,n}^S$ and  $\boldsymbol{X}_{m,n}^L$, since $\boldsymbol{X}_{m,n}^S$ and  $\boldsymbol{X}_{m,n}^L$ can be determined using $\mathcal{B}_{m,n}^S$ and $\mathcal{B}_{m,n}^L$, respectively.
{\color{black}$\boldsymbol{X}_{m,n}^S$ and $\boldsymbol{X}_{m,n}^L$ as the variables of problem (11) cooperatively construct the completed description, that is used by the listener for decision-making. The optimization problem (11) is solved if the speaker and listener can determine the transmitted and inferred descriptions in a way that the resulting completed description is the perfect and abstract description of each event. Therefore, solving problem (11) is equivalent to finding the perfect and abstract description of each event by the pair of the speaker and listener.}
\eqref{Equation:Constraint_3} indicates that a belief can only be transmitted by the speaker or can be inferred by the listener. \eqref{Equation:Constraint_4} indicates that, at each time slot, only one belief from each level of the hierarchy can be used. \eqref{Equation:Constraint_5} captures the fact that the existence of a belief from level $k$ is necessary for the presence of a belief from level $k+1$.

{\color{black}Since the environment's dynamic nature affects the observed event's transition, the speaker and listener need to interact with the environment to experience different events and descriptions. We first note that classical stochastic optimization techniques cannot be used to solve (11) because they require knowing the transition functions of the system, which are unknown. Second, model-based RL methods also cannot be used because the speaker and listener are unaware of the transition model of the environment (matrices $\boldsymbol{P}$ and $\boldsymbol{\Tilde{P}}$) \cite{sutton2018reinforcement}. Finally, model-free methods will not be efficient due to the specific characteristics of our problem including (a) the massive size of action space, $2^B$, which is exponential in the number of beliefs, and (b) the task execution structure of the problem that exhibits a sparse reward signal.}

These challenges of using classical RL motivate us to introduce a curriculum framework \cite{JMLR:v21:20-212} combined with multi-agent RL to solve problem \eqref{Equation:Optimization}.
The main goal of the speaker and listener is to cooperatively find the perfect and abstract description of each event to solve \eqref{Equation:Optimization}. Here, the main responsibility of the speaker is finding useful beliefs for the description of each event, and the listener wants to find useful beliefs to complete the speaker's transmitted description. However, the speaker and listener are completely blind about the usefulness of each belief for the different events, and, thus, they must gain this knowledge by experiencing various tasks with different events. However, the random exploration process of classical RL requires a long training to converge because of the massive action space size. In such problems, using a curriculum framework can
be a suitable solution. The main idea behind CL is to design simple tasks based on a main problem which requires long training, perform these simple
tasks, and use their gained experience to solve the main problem more efficiently and more quickly \cite{JMLR:v21:20-212}. 
Here, we propose a CL method that enables the speaker and listener to identify, step-by-step, the hierarchical structure of the belief set. In each step, the speaker and listener identify a portion of the hierarchical structure and learn the perfect description of the events that can be described using the identified portion of the belief set. Note that the identified portion of the belief set in each step is used to initialize the algorithm in the next step.
\vspace{-5.5mm}
\subsection{Bottom-to-Up Curriculum Learning Algorithm}
\vspace{-2.5mm}
The hierarchical structure of the belief set plays a key role in determining the perfect description of each event.
However, the speaker and listener are unaware of the exact hierarchical structure of the belief set.
An interesting solution here is to gradually identify the hierarchical structure of the belief set and the perfect descriptions of the events that can be perfectly described using the identified portion of the hierarchical structure. 
We thus propose a linear sequence-based and task-level CL, defined by a sequence of problems $[z_1, z_2, \ldots, z_l]$ \cite{JMLR:v21:20-212}. Here, $z_l$ refers to the problem that should be solved in CL step $l$ and whose gained knowledge will then be used to initialize and solve $z_{l+1}$.
The term “task” is used here in the context of the CL paradigm, and it is not related to our system’s task. In $z_l$, the CL method determines the structure of the first $l+1$ levels of the belief set hierarchy as well as the perfect descriptors of events using these $l+1$ levels of the belief set. 
We consider two outputs for each CL step. The first one is related to the perfect description of each event and the second is related to the hierarchical belief structure. Note that the minimum number of beliefs for having a perfect description of an event is two. Thus, during the first CL step, the speaker and listener will learn the structure of the first and second hierarchy levels. 
Let $\mathcal{B}^{\text{Out}}_{z_1, e} = \Big\{\big(b_{i_1}, b_{i_2}\big) \Big| b_{i_1} \in \mathcal{B}_1, b_{i_2} \in \mathcal{B}_2, \, \big(b_{i_1}, b_{i_2}\big) \in \mathcal{B}_{e}^P\Big\}$, be the output of the first CL step for the perfect description of event $e$,
where $\mathcal{B}_1$ and $\mathcal{B}_2$ are the belief sets of the first and second hierarchy levels. For some event $e^{\prime}$, we may have $\mathcal{B}^{\text{Out}}_{z_1, e^{\prime}} = \emptyset$, 
if this event does not have a perfect descriptor with 2 beliefs. In other words, the first CL step determines the perfect description of events that have $\mathcal{B}_h \in \mathcal{B}_{e}^P$ with $K_h = 2$. 
Now, let $B^{\text{Out}}_{z1,H}$ be the second output of $z_1$ that is related to the hierarchical structure of the belief set, as follows:
\vspace{-3.8mm}
\begin{align}
    \mathcal{B}^{\text{Out}}_{z_1,H} = \left\{\big(b_{i_1}, b_{i_2}\big) \Big| \exists e \in \mathcal{E}: \big(b_{i_1}, b_{i_2}\big) \in \mathcal{B}^{\text{Out}}_{z_1, e}\right\}.
    \label{Equation:CL_Out_1_Hierarchy}
\end{align}

\begin{figure*}[!t]
\centering
\begin{tikzpicture}[scale=0.65, every node/.style={scale=0.5}]

\draw[black, very thick] (0.5,2) rectangle (3.5,0);
\draw[black, very thick] (5,2) rectangle (9,-1);
\draw[black, very thick] (11,2) rectangle (16,-2);

\draw [->, color=black, thick] (3.5,1) -- (5,1);
\draw [->, color=black, thick] (9,1) -- (11,1);

\node[text width = 6cm] at (2.5,2.3) {1st step of CL determines $\mathcal{B}^{\text{Out}}_{z_1,H}$};
\node[text width = 6cm] at (7.5,2.3) {2nd step of CL determines $\mathcal{B}^{\text{Out}}_{z_2,H}$};
\node[text width = 6cm] at (13.8,2.3) {3rd step of CL determines $\mathcal{B}^{\text{Out}}_{z_3,H}$};


\draw[color=blue](1.5,1.5) circle (0.25);
\draw[color=blue](2.5,1.5) circle (0.25);

\draw[color=blue](1,0.5) circle (0.25);
\draw[color=blue](2,0.5) circle (0.25);
\draw[color=blue](3,0.5) circle (0.25);

\draw [->, color=blue, thick] (1.5,1.25) -- (1,0.75);
\draw [->, color=blue, thick] (1.5,1.25) -- (2,0.75);
\draw [->, color=blue, thick] (1.5,1.25) -- (3,0.75);

\draw [->, color=blue, thick] (2.5,1.25) -- (2,0.75);
\draw [->, color=blue, thick] (2.5,1.25) -- (3,0.75);

\draw[color=blue](6.5,1.5) circle (0.25);
\draw[color=blue](7.5,1.5) circle (0.25);

\draw[color=blue](6,0.5) circle (0.25);
\draw[color=blue](7,0.5) circle (0.25);
\draw[color=blue](8,0.5) circle (0.25);

\draw[color=green](5.5,-0.5) circle (0.25);
\draw[color=green](6.5,-0.5) circle (0.25);
\draw[color=green](7.5,-0.5) circle (0.25);
\draw[color=green](8.5,-0.5) circle (0.25);

\draw [->, color=blue, thick] (6.5,1.25) -- (6,0.75);
\draw [->, color=blue, thick] (6.5,1.25) -- (7,0.75);
\draw [->, color=blue, thick] (6.5,1.25) -- (8,0.75);

\draw [->, color=blue, thick] (7.5,1.25) -- (7,0.75);
\draw [->, color=blue, thick] (7.5,1.25) -- (8,0.75);

\draw [->, color=green, thick] (6,0.25) -- (5.5,-0.25);
\draw [->, color=green, thick] (6,0.25) -- (6.5,-0.25);

\draw [->, color=green, thick] (7,0.25) -- (5.5,-0.25);
\draw [->, color=green, thick] (7,0.25) -- (6.5,-0.25);
\draw [->, color=green, thick] (7,0.25) -- (7.5,-0.25);
\draw [->, color=green, thick] (7,0.25) -- (8.5,-0.25);

\draw [->, color=green, thick] (8,0.25) -- (7.5,-0.25);
\draw [->, color=green, thick] (8,0.25) -- (8.5,-0.25);

\draw[color=blue](13,1.5) circle (0.25);
\draw[color=blue](14,1.5) circle (0.25);

\draw[color=blue](12.5,0.5) circle (0.25);
\draw[color=blue](13.5,0.5) circle (0.25);
\draw[color=blue](14.5,0.5) circle (0.25);

\draw[color=green](12,-0.5) circle (0.25);
\draw[color=green](13,-0.5) circle (0.25);
\draw[color=green](14,-0.5) circle (0.25);
\draw[color=green](15,-0.5) circle (0.25);

\draw[color=red](11.5,-1.5) circle (0.25);
\draw[color=red](12.5,-1.5) circle (0.25);
\draw[color=red](13.5,-1.5) circle (0.25);
\draw[color=red](14.5,-1.5) circle (0.25);
\draw[color=red](15.5,-1.5) circle (0.25);

\draw [->, color=blue, thick] (13,1.25) -- (12.5,0.75);
\draw [->, color=blue, thick] (13,1.25) -- (13.5,0.75);
\draw [->, color=blue, thick] (13,1.25) -- (14.5,0.75);

\draw [->, color=blue, thick] (14,1.25) -- (13.5,0.75);
\draw [->, color=blue, thick] (14,1.25) -- (14.5,0.75);

\draw [->, color=green, thick] (12.5,0.25) -- (12,-0.25);
\draw [->, color=green, thick] (12.5,0.25) -- (13,-0.25);

\draw [->, color=green, thick] (13.5,0.25) -- (12,-0.25);
\draw [->, color=green, thick] (13.5,0.25) -- (13,-0.25);
\draw [->, color=green, thick] (13.5,0.25) -- (14,-0.25);
\draw [->, color=green, thick] (13.5,0.25) -- (15,-0.25);

\draw [->, color=green, thick] (14.5,0.25) -- (14,-0.25);
\draw [->, color=green, thick] (14.5,0.25) -- (15,-0.25);

\draw [->, color=red, thick] (12,-0.75) -- (11.5,-1.25);
\draw [->, color=red, thick] (12,-0.75) -- (12.5,-1.25);

\draw [->, color=red, thick] (13,-0.75) -- (11.5,-1.25);
\draw [->, color=red, thick] (13,-0.75) -- (12.5,-1.25);
\draw [->, color=red, thick] (13,-0.75) -- (13.5,-1.25);

\draw [->, color=red, thick] (14,-0.75) -- (13.5,-1.25);
\draw [->, color=red, thick] (14,-0.75) -- (14.5,-1.25);
\draw [->, color=red, thick] (14,-0.75) -- (15.5,-1.25);

\draw [->, color=red, thick] (15,-0.75) -- (14.5,-1.25);
\draw [->, color=red, thick] (15,-0.75) -- (15.5,-1.25);

\end{tikzpicture}
\vspace{-5mm}
\caption{\small{An example of how the proposed CL method determines the structure of the belief set.}}
\label{Fig:CL_Flowchart}
\vspace{-10mm}
\end{figure*}
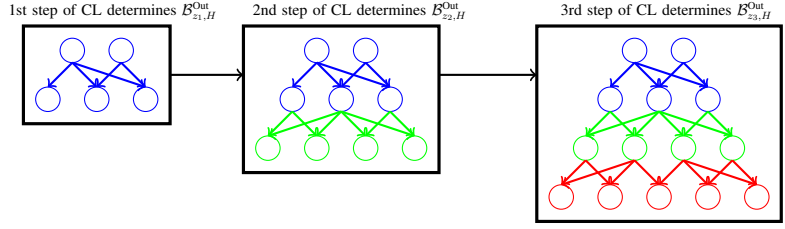

In the second CL step $z_2$, the speaker and listener want to find the beliefs in the third level of the belief set hierarchy while also determining the compatible combination between the beliefs in the first and second level of the belief set hierarchy and the beliefs in the third level of the hierarchy. For this purpose, they use the output of the first CL step as the compatible combinations between the first and second levels of the hierarchy.
Let $\mathcal{B}^{\text{Out}}_{z_2, e}$ be the output of the second CL step that is related to possible perfect descriptions of the events and defined as follows:
\vspace{-3.7mm}
\begin{align}
    \mathcal{B}^{\text{Out}}_{z_2, e} = \left\{\big(b_{i_1}, b_{i_2}, b_{i_3}\big) \Big| \big(b_{i_1}, b_{i_2}\big) \in \mathcal{B}^{\text{Out}}_{z_1,H}, b_{i_3} \in \mathcal{B}_3, \big(b_{i_1}, b_{i_2}, b_{i_3}\big) \in \mathcal{B}_{e}^P\right\},
    \label{Equation:CL_Out_2_Event}
\end{align}
where $\mathcal{B}^{\text{Out}}_{z_1,H}$ is defined in \eqref{Equation:CL_Out_1_Hierarchy} and $\mathcal{B}_3$ is the belief set of the third level of the hierarchy. As seen in \eqref{Equation:CL_Out_2_Event}, the gained knowledge $\mathcal{B}^{\text{Out}}_{z_1,H}$, at the first step is directly used for
initializing $z_2$, which showcases how we leverage the CL concept in our approach. Let $\mathcal{B}^{\text{Out}}_{z_2, H}$ be the second output of $z_2$, which is related to the hierarchical structure of the belief set, which is $\mathcal{B}^{\text{Out}}_{z_2,H} = \left\{\big(b_{i_1}, b_{i_2}, b_{i_3}\big) \Big| \exists e \in \mathcal{E}: \big(b_{i_1}, b_{i_2}, b_{i_3}\big) \in \mathcal{B}^{\text{Out}}_{z_2, e}\right\}$.
By following the procedure used for defining the outputs of $z_1$ and $z_2$, we can determine the two outputs of CL step $z_l$ for $l \leq K-1$, as follows:
\vspace{-11mm}
\begin{align}
    \mathcal{B}^{\text{Out}}_{z_l, e} &= \Big\{\big(b_{i_1}, \ldots, b_{i_{l+1}}\big) \Big| \big(b_{i_1}, \ldots, b_{i_l}\big) \in \mathcal{B}^{\text{Out}}_{z_{l-1},H}, b_{i_{l+1}} \in \mathcal{B}_{l+1}, \big(b_{i_1}, \ldots, b_{i_{l+1}}\big) \in \mathcal{B}_{e}^P\Big\}. 
    \label{Equation:CL_Out_l_Event} \\
    \mathcal{B}^{\text{Out}}_{z_l,H} &= \Big\{\big(b_{i_1}, \ldots, b_{i_{l+1}}\big) \Big| \exists e \in \mathcal{E}: \big(b_{i_1}, \ldots, b_{i_{l+1}}\big) \in \mathcal{B}^{\text{Out}}_{z_l, e} \Big\},
    \label{Equation:CL_Out_l_Hierarchy} 
\end{align}
where $\mathcal{B}^{\text{Out}}_{z_l,H}$ is the set of compatible combinations between the beliefs in the first $l$ levels of the hierarchy and the beliefs in level $l+1$. Also, $\mathcal{B}^{\text{Out}}_{z_l, e}$ is the set of perfect descriptors of event $e$ that have $(l+1)$ beliefs. 
{\color{black}The main objective of step $l$ of the CL algorithm is determining $\mathcal{B}^{\text{Out}}_{z_l, e}$ and $\mathcal{B}^{\text{Out}}_{z_l, H}$. For this purpose, we use the output of the CL algorithm at step $l-1$, $\mathcal{B}^{\text{Out}}_{z_{l-1}, H}$, which is the hierarchical structure of the first $l$ levels of the belief set.}
The CL method proceeds until we have $l=K-1$, at which point the algorithm determines the compatible combinations in the last belief hierarchy level and the $K$-beliefs perfect descriptors of each event $e$. 
Fig. \ref{Fig:CL_Flowchart} shows an example of the proposed algorithm for determining the structure of a belief set that has $4$ belief levels. Fig. \ref{Fig:CL_Flowchart} shows how the output $\mathcal{B}^{\text{Out}}_{z_l,H}$ of each step of the CL is used in the next step.

\vspace{-7mm}
\subsection{Reinforcement Learning at each CL Step}
\vspace{-2mm}
For solving each CL step, we now we introduce a multi-agent RL method (the agents are the speaker and listener) using Q-Learning. 
{\color{black} In each step of CL, a multi-agent RL problem is defined. As the CL algorithm evolves, the defined multi-agent RL problem in each CL step becomes more difficult to solve. However, our CL approach initializes the multi-agent RL problem of each CL step based on the output of the previous step thus reducing this difficulty and making the multi-agent RL problem tractable. The CL method continues until the main problem is completely and tractably solved.}
For our system, in each event, we want to find a single perfect description, which minimizes the incurred cost of the speaker and listener, minimizes task execution time, and maximizes the belief efficiency.
The speaker and listener will cooperate to solve our multi-agent RL problem, in each step. 
Here, the speaker will learn how to transmit an abstract description, i.e., a subset of the perfect description of the observed event. Then, the listener learns how to infer the beliefs based on the speaker's transmitted description such that the completed description is a perfect description of the observed event.
In the first CL step, $z_1$, the speaker and listener are unaware of the hierarchical structure of the belief set. In $z_l$, where $l \geq 2$, the speaker and listener are aware of the first $l$ hierarchy levels. 
Hence, we first introduce the learning method for the first CL step. Then, we present the learning method for any step $l \geq 2$ of the CL method.

\subsubsection{Learning during the First CL Step}
In the first CL step, we define the state space and action space of the speaker as follows:
\vspace{-3mm}
\begin{align}
    \Omega_{\Lambda, z_1}^{S} = \Big\{e \Big| e \in \mathcal{E} \Big\}, \quad
    \Omega_{A, z_1}^{S} = \Big\{b_i \Big| b_i \in \mathcal{B} \Big\},
    \label{Equation:CL_Speaker_1_State_Action}
\end{align}
where $\Omega_{\Lambda, z_1}^{S}$ and $\Omega_{A, z_1}^{S}$ are the state space and action space of the speaker in the first CL step. The action space is all possible
subsets of $\mathcal{B}$ with cardinality $1$. We define $\lambda_{m, n, z_1}^S \in \Omega_{\Lambda, z_1}^{S}$ and $a_{m, n, z_1}^S \in \Omega_{A, z_1}^{S}$ as the state and action of the speaker at slot $n$ of episode $m$, in the first CL step.
The state space and action spaces of the listener in the first CL step are defined as follows:
\vspace{-2mm}
\begin{align}
    \Omega_{\Lambda, z_1}^{L} = \Big\{b_i \Big| b_i \in \mathcal{B} \Big\}, \quad
    \Omega_{A, z_1}^{L}\big(\lambda_{m, n, z_1}^L\big) = \Big\{b_i \Big| b_i \in \big\{\mathcal{B} \setminus \lambda_{m, n, z_1}^L\big\} \Big\},
    \label{Equation:CL_Listener_1_State_Action}
\end{align}
where $\Omega_{\Lambda, z_1}^{L}$ is the state space of the listener in the first step of CL and $\lambda_{m, n, z_1}^L \in \Omega_{\Lambda, z_1}^{L}$ represents the state of the listener at slot $n$ of episode $m$, in the CL first step. Also, $\Omega_{A, z_1}^{L}\big(\lambda_{m, n, z_1}^L\big)$ is the action space for state $\lambda_{m, n, z_1}^L$. For the listener, we should define a specific action space for each state in the CL first step. For state $\lambda_{m, n, z_1}^L$, the action space is all possible subsets of $\{\mathcal{B} \setminus \lambda_{m, n, z_1}^L\}$ with cardinality $1$.
We define $a_{m, n, z_1}^L \in \Omega_{A, z_1}^{L}\big(\lambda_{m, n, z_1}^L\big)$ as the action of the listener at slot $n$ of episode $m$, in the first CL step. We can now connect the state and action of the speaker and listener to the variables of the original problem \eqref{Equation:Optimization}, as follows:
\vspace{-4mm}
\begin{alignat}{3}
    \lambda_{m, n, z_1}^S &= e_{m,n}^S,  \quad \mathcal{B}_{m,n}^S &&= a_{m, n, z_1}^S, 
    \label{Equation:CL_Speaker_1_Param}\\
    \lambda_{m, n, z_1}^L &= \mathcal{B}_{m,n}^S, \quad \mathcal{B}_{m,n}^L &&= a_{m, n, z_1}^L,  
    \label{Equation:CL_Listener_1_Param}
\end{alignat}
where $e_{m,n}^S$ is the observed event of the speaker. $\mathcal{B}_{m,n}^S$ and $\mathcal{B}_{m,n}^L$ are the transmitted and inferred description of the speaker and listener, respectively. As seen in \eqref{Equation:CL_Listener_1_Param}, the RL state of the listener is the transmitted description of the speaker because the listener is not aware of the observed event and, thus, it should make decisions based on the speaker's transmitted description. The completed description, $\mathcal{B}_{m,n}$, used by the listener for reconstructing the observed event, will be $\mathcal{B}_{m,n} = \left\{\mathcal{B}_{m,n}^S, \mathcal{B}_{m,n}^L \right\} = \left\{a_{m,n,z_1}^S, a_{m,n,z_1}^L \right\}$.

The \emph{reward function} of the speaker at the first CL step is $R_{z_1}^S:\Omega_{\Lambda, z_1}^{S} \times \Omega_{A, z_1}^{S} \times \Omega_{\Lambda, z_1}^{L} \times \Omega_{A, z_1}^{L} \to \Omega_{R, z_1}^{S}$, and it captures the expected immediate reward gained by the speaker for taking each action in each state, which is also dependent on the state and action taken by the listener.
We define $R_{z_1}^S(\lambda_{m,n,z_1}^S, a_{m,n,z_1}^S, \lambda_{m,n,z_1}^L, a_{m,n,z_1}^L)$ as the reward gained by the speaker for taking action $a_{m,n,z_1}^S$ at state $\lambda_{m,n,z_1}^S$, while the listener takes action $a_{m,n,z_1}^L$ at state $\lambda_{m,n,z_1}^L$.
To define the reward, we recall that the speaker's goal is to determine the perfect description of each event that simultaneously optimizes the belief transmission cost, task execution time, and belief efficiency. Thus, we should consider how each action can affect the aforementioned metrics for the speaker. The speaker's incurred cost for taken action and the effect of the taken action on the belief efficiency can be easily derived. However, the effect of each action on the task execution procedure is determined based on the current and next observed events. Thus, we have:
\vspace{-3.5mm}
\begin{align}
    R_{z_1}^S\big(\lambda_{m,n,z_1}^S, a_{m,n,z_1}^S, \lambda_{m,n,z_1}^L, a_{m,n,z_1}^L\big) &= \begin{cases} - C_{m,n}^S - W_{m,n}^S + R_T, & \lambda_{m,n+1,z_1}^S \in \mathcal{E}_{\text{fin}},   \\ 
								      -C_{m,n}^S -W_{m,n}^S - C_T, & \lambda_{m,n+1,z_1}^S \in \mathcal{E}_{\text{init}}, \\
								      -C_{m,n}^S -W_{m,n}^S, & \lambda_{m,n+1,z_1}^S \in \mathcal{E}_{\text{mid}},  \end{cases}
\label{Equation:Reward_Def_Speaker_First_Step} 
\end{align}
where $C_{m,n}^S$ and $W_{m,n}^S$ are defined in \eqref{Equation:Slot_Cost_Speaker_Listener} and \eqref{Equation:Episode_Belief_Efficiency}, respectively. $T$ is the executed task in episode $m$. $R_T$ is the reward of executing task $T$ and $C_T$ is the cost of delaying task $T$. $\lambda_{m,n+1,z_1}^S$ is the speaker's next state, which is equal to the observed event of the speaker at slot $n+1$, which is determined using \eqref{Equation:Transition_Prob}.
The reward function in \eqref{Equation:Reward_Def_Speaker_First_Step} is defined in a way that solve \eqref{Equation:Optimization}. Specifically, $C_{m,n}^S$, which contributes to the objective function of \eqref{Equation:Optimization}, is considered as a cost in the reward function to minimize the transmission cost of the speaker. $W_{m,n}^S$ which contributes to \eqref{Equation:Constraint_Belief} is considered as a cost to maximize the belief efficiency according to \eqref{Equation:Episode_Belief_Efficiency}. Finally, $R_T$ as the reward of executing task $T$, and $C_T$ as the cost of delaying task $T$ are used in the RL reward function to satisfy the task execution time constraint in  \eqref{Equation:Constraint_Time}.

Similarly, the listener's \emph{reward function} at the first CL step is defined as $R_{z_1}^L:\Omega_{\Lambda, z_1}^{S} \times \Omega_{A, z_1}^{S} \times \Omega_{\Lambda, z_1}^{L} \times \Omega_{A, z_1}^{L} \to \Omega_{R, z_1}^{L}$, and it captures the immediate gained reward of the listener for taking each action in each state. The listener's reward function is also dependent on the state and the action of the speaker, and will capture the cost of each action as well as how each action in each state affects the task execution procedure. We assume that the speaker informs the listener about the observed event type (initial, intermediary, or final) at each slot. Hence, we have:
\vspace{-3.5mm}
\begin{align}
    R_{z_1}^L\big(\lambda_{m,n,z_1}^S, a_{m,n,z_1}^S, \lambda_{m,n,z_1}^L, a_{m,n,z_1}^L\big) &= \begin{cases} - C_{m,n}^L + R_T, & \lambda_{m,n+1,z_1}^S \in \mathcal{E}_{\text{fin}},   \\ 
								      -C_{m,n}^L - C_T, & \lambda_{m,n+1,z_1}^S \in \mathcal{E}_{\text{init}}, \\
								      -C_{m,n}^L, & \lambda_{m,n+1,z_1}^S \in \mathcal{E}_{\text{mid}}  ,\end{cases}
\label{Equation:Reward_Def_Listener_First_Step} 
\end{align}
where $C_{m,n}^L$ is the listener's incurred cost for the inferred description at each slot, as per \eqref{Equation:Slot_Cost_Speaker_Listener}. Similar to speaker's case, the reward function in \eqref{Equation:Reward_Def_Listener_First_Step} is defined in a way to solve problem \eqref{Equation:Optimization}. 
{\color{black}The completed description in each time slot, $\mathcal{B}_{m,n}$, can include multiple beliefs from a single level. However, such a description cannot perfectly describe an event. Therefore, the reward of such descriptions would be calculated using the second or third case of (20) and (21). On the other hand, the reward of the perfect descriptions would be calculated using the first or second case of (20) and (21).
Hence, the value of the perfect descriptions in each state at the end of the learning time will be greater than the imperfect descriptions.}
\subsubsection{Learning at CL Step $l$}
For CL step $l \geq 2$, the state and action spaces are as follows:
\vspace{-4mm}
\begin{align}
    \Omega_{\Lambda, z_l}^{S} = \Big\{e \Big| e \in \mathcal{E} \Big\}, \quad
    \Omega_{A, z_l}^{S} = \Big\{\big(b_{i_1}, \ldots, b_{i_l} \big) \Big| \big(b_{i_1}, \ldots, b_{i_l} \big) \in \mathcal{B}^{\text{Out}}_{z_{l-1},H} \Big\}, 
    \label{Equation:CL_Speaker_l_State_Action}
\end{align}
where $\Omega_{\Lambda, z_l}^{S}$ and $\Omega_{A, z_l}^{S}$ are the state space and action space of the speaker in CL step $l$. We define $\lambda_{m, n, z_l}^S \in \Omega_{\Lambda, z_l}^{S}$ and $a_{m, n, z_l}^S \in \Omega_{A, z_l}^{S}$ as the state and action of the speaker at slot $n$ of episode $m$, in CL step $l$. According to this definition, the state space is the same as the first CL step. However, we define the action space based on the output of the previous CL step. More precisely, in CL step $l$, the speaker can only use the compatible combinations between the beliefs, $\mathcal{B}^{\text{Out}}_{z_{l-1},H}$, in the first $l$ levels of the hierarchical structure of the belief set for the description of the observed events. Now, we define the state space and action space of the listener in CL step $l$, as follows:
\vspace{-4.4mm}
\begin{align}
    \Omega_{\Lambda, z_l}^{L} &= \Big\{\big(b_{i_1}, \ldots, b_{i_l} \big) \Big| \big(b_{i_1}, \ldots, b_{i_l} \big) \in \mathcal{B}^{\text{Out}}_{z_{l-1},H} \Big\}, \label{Equation:CL_Listener_l_State} \\
    \Omega_{A, z_l}^{L} &= \Big\{b_i \Big| b_i \in \big\{\mathcal{B}  \setminus \bigcup_{k=1}^{l}{\mathcal{B}_k}\big\} \Big\}, \quad \bigcup_{k=1}^{l}{\mathcal{B}_k} = \bigcup_{\mathcal{B}_{l^{\prime}} \in \mathcal{B}^{\text{Out}}_{z_{l-1},H}}{\mathcal{B}_{l^{\prime}}},
    \label{Equation:CL_Listener_l_Action}
\end{align}
where $\Omega_{\Lambda, z_l}^{L}$ and $\Omega_{A, z_l}^{L}$ are the state and action spaces of the listener at step $l$ of the CL method. We define $\lambda_{m, n, z_l}^L \in \Omega_{\Lambda, z_l}^{L}$ and $a_{m, n, z_l}^L \in \Omega_{A, z_l}^{L}$ as the state and action of the listener at slot $n$ of episode $m$. For defining the action set of the listener at CL step $l \geq 2$, 
we consider the fact that the speaker and listener are aware of the first $l$ levels of the hierarchical structure of the belief set, based on the output of CL step $(l-1)$. We have the same connection between the state and action of the speaker and listener to the variables of the original problem at each time slot of step $l$ of the CL, as defined in \eqref{Equation:CL_Speaker_1_Param} and \eqref{Equation:CL_Listener_1_Param}. Also, the reward functions of the speaker and listener are defined in a similar way to the first step, as per \eqref{Equation:Reward_Def_Speaker_First_Step} and \eqref{Equation:Reward_Def_Listener_First_Step}. 

\begin{algorithm}[t]
	\caption{\small{Bottom-Up Curriculum Learning Algorithm}}
	\label{Algo_CL}
\footnotesize{
\vspace{-1mm}
\textbf{Inputs:} $\mathcal{E}, \mathcal{B},  C_{k,u}^S$: The cost of transmitting belief $u$ of level $k$ for the speaker, $C_{k,u}^L$: The cost of inferring belief $u$ of level $k$ for the listener. $C_T$: The cost of delaying task $T$, and $R_T$: The cost of executing task $T$. \label{Algo:Input1}\\\vspace{-1mm}
$\mathcal{B}_{\text{Remain}} = \mathcal{B}$ and $\mathcal{B}_{e}^P = \{\}, \; \forall e \in \mathcal{E}$.\label{Algo:Initial1}\\
    \vspace{-1mm}
	\While{($\mathcal{B}_{\text{Remain}} \neq \emptyset$)} 
	{\label{Algo_CL_CL_Start}\vspace{-2mm}
		\If{$ l = 1$}
		{\label{Algo_CL_L1_Start}
		    \vspace{-2mm}
			Compute $\Omega_{\Lambda,z_1}^S$ and $\Omega_{A,z_1}^S$ using \eqref{Equation:CL_Speaker_1_State_Action}. \label{Algo:CL_Speaker_1_State_Action} and 
			 $\Omega_{\Lambda,z_1}^L$ and $\Omega_{A,z_1}^L$ using \eqref{Equation:CL_Listener_1_State_Action}. \label{Algo:CL_Listener_1_State_Action}\vspace{-1mm}\\
			Solve a multi-agent RL including speaker and listener as the two agents, using $\Omega_{\Lambda,z_1}^S$, $\Omega_{\Lambda,z_1}^L$, $\Omega_{A,z_1}^S$, and $\Omega_{A,z_1}^S$ as the state and action spaces, and \eqref{Equation:Reward_Def_Speaker_First_Step} and \eqref{Equation:Reward_Def_Listener_First_Step} as the reward functions. \label{Algo:Multi_Agent_RL_First_Step}\vspace{-1mm}\\
			Compute $\mathcal{B}^{\text{Out}}_{z_1, e}$ for each $e \in \mathcal{E}$, using $Q$ values and $\mathcal{B}_{e}^P = \mathcal{B}_{e}^P \bigcup \mathcal{B}^{\text{Out}}_{z_1, e}$. 
			\label{Alog:CL_Out_1_Event}\vspace{-1mm}\\
			Compute $\mathcal{B}^{\text{Out}}_{z_1, H}$ using \eqref{Equation:CL_Out_1_Hierarchy} and  $\mathcal{B}_{\text{Remain}} = \mathcal{B}_{\text{Remain}} \setminus \mathcal{B}^{\text{Out}}_{z_1, H}$ \label{Alog:CL_Out_1_Hierarchy}.
			\vspace{-3mm}
		}\label{Algo_CL_L1_End}
		\vspace{-2mm}
		\If{$ l > 1$}
		{\label{Algo_CL_L2_Start}
		    \vspace{-2mm}
			Compute $\Omega_{\Lambda,z_l}^S$ and $\Omega_{A,z_l}^S$ using \eqref{Equation:CL_Speaker_l_State_Action}.\vspace{-1mm}\\
			Compute $\Omega_{\Lambda,z_l}^L$ using \eqref{Equation:CL_Listener_l_State} and $\Omega_{A,z_l}^L$ using \eqref{Equation:CL_Listener_l_Action}.\vspace{-1mm}\\
			Solve a multi-agent RL including speaker and listener as the two agents, using $\Omega_{\Lambda,z_l}^S$, $\Omega_{\Lambda,z_l}^L$, $\Omega_{A,z_l}^S$, and $\Omega_{A,z_l}^S$ as the state and action spaces, and \eqref{Equation:Reward_Def_Speaker_First_Step} and \eqref{Equation:Reward_Def_Listener_First_Step} as the reward functions.\vspace{-1mm}\\
			Compute $\mathcal{B}^{\text{Out}}_{z_l, e}$ for each $e$, using $Q$ values and $\mathcal{B}_{e}^P = \mathcal{B}_{e}^P \bigcup \mathcal{B}^{\text{Out}}_{z_l, e}$.\vspace{-1mm}\\
			Compute $\mathcal{B}^{\text{Out}}_{z_l, H}$ using \eqref{Equation:CL_Out_l_Hierarchy} and  $\mathcal{B}_{\text{Remain}} = \mathcal{B}_{\text{Remain}} \setminus \mathcal{B}^{\text{Out}}_{z_l, H}$.
			\vspace{-3mm}
		}\label{Algo_CL_L2_End}
		\vspace{-3mm}
	}\label{Algo_CL_CL_End}
\vspace{-2mm}
\textbf{Outputs:} $\mathcal{B}^P_{e}, \; \forall e \in \mathcal{E}$.\label{Algo_CL_Output_Cal} 
}
\end{algorithm}
\vspace{-6.5mm}
\subsection{Proposed Curriculum Learning Algorithm Review}
\vspace{-2.5mm}
{\color{black}Algorithm 1 summarizes our CL framework for solving (11). In this algorithm, in each step of the CL method, a multi-agent RL problem which is initialized based on the output of the previous step of CL is solved. At the end of CL, the structure of the hierarchical belief set and the abstract and perfect description of each event are determined, which is the solution of (11). To our best knowledge, this algorithm is \emph{the first implementation of a CL-based reinforcement learning process} that can be used to solve semantic communication problems. Our CL method includes $K-1$ steps, where the value of $K$ is unknown at the beginning of CL.}
In step \ref{Algo:Input1}, we determine the input variables of the algorithm. In step \ref{Algo:Initial1}, we initialize two main parameters of the algorithm, where $\mathcal{B}_{\text{Remain}}$ is the set of beliefs whose levels in the hierarchical structure are not yet determined. $\mathcal{B}_{e}^P$ is the set of all possible perfect descriptors of event $e$. In steps \ref{Algo_CL_L1_Start}-\ref{Algo_CL_L1_End}, 
we show the first CL step, while in steps \ref{Algo_CL_L2_Start}-\ref{Algo_CL_L2_End} we show CL steps $l \geq 2$. For the first CL step, the state and action spaces of the speaker and listener are determined in step \ref{Algo:CL_Speaker_1_State_Action}. In step \ref{Algo:Multi_Agent_RL_First_Step}, the multi-agent RL problem of the first CL step is performed. In step \ref{Alog:CL_Out_1_Event}, the first output of the CL method which is related to event description is calculated and the value of $\mathcal{B}_{e}^P$ is updated. Then, in step \ref{Alog:CL_Out_1_Hierarchy}, the second output of the CL method related to the hierarchical structure of the belief set is calculated and the value of $\mathcal{B}_{\text{Remain}}$ is updated. We have the same procedure for steps $l \geq 2$ of the CL methods, which are captured by steps \ref{Algo_CL_L2_Start}-\ref{Algo_CL_L2_End} of Algorithm \ref{Algo_CL}. The CL output is the set of all possible perfect descriptors of event $e$, captured by $\mathcal{B}_{e}^P$.
\vspace{-6mm}
\subsection{Algorithm Analysis}
\vspace{-2mm}
The main output of problem \eqref{Equation:Optimization} is determining the perfect description of each event. Next, we derive a sufficient condition for guaranteeing that our CL method can determine the possible perfect descriptors of each event, at the end of each step of the CL when the multi-agent RL problem of that step converges\footnote{Note that deriving a necessary and sufficient condition here is analytically challenging.}. Note that, it is completely possible that for some events $e \in \mathcal{E}$, we have $\mathcal{B}^{\text{Out}}_{z_l, e} = \emptyset$, where $\mathcal{B}^{\text{Out}}_{z_l, e}$ is the set of $l$-beliefs perfect descriptors of $e$. This because, for such events, there is no $l$-beliefs perfect descriptor.
\vspace{-4mm}
\begin{thm}\label{thm1}
A sufficient condition for finding the possible perfect descriptors of each event at the end of each step of the proposed CL method is that the reward, $R_T$, satisfies the following inequality:
\vspace{-6mm}
\begin{align}
R_T &\geq \max\Big(D_1, D_2\Big), \label{Equation_Constraint_Total}\\
D_1 &= \max_{T \in \mathcal{T}}{\bigg(\dfrac{\max{\big(\Delta^S, \Delta^L\big)}}{\min_{1 \leq j \leq L_T^{\text{max}}-1}{\big[\boldsymbol{P}(j, L_T^{\text{max}}) + \boldsymbol{\Tilde{P}}(j,1)\big]}}\bigg)}, \label{Equation_Constraint_Inside_Final}\\
D_2 &= \max_{T \in \mathcal{T}}{\bigg(\dfrac{\max{(\Delta^S, \Delta^L)}}{\min_{2 \leq j \leq L_T^{\text{max}}-1}{\big[\boldsymbol{P}(j, L_T^{\text{max}}) - \boldsymbol{P}(j-1, L_T^{\text{max}})\big]}}\bigg)}, \label{Equation_Constraint_Outside_Final}\\
\Delta^S &= \max_{\mathcal{B}_1, \mathcal{B}_2 \subset \mathcal{B}}{\Big|C^S(\mathcal{B}_1) - C^S(\mathcal{B}_2)}\Big|, \quad 
\Delta^L = \max_{\mathcal{B}_1, \mathcal{B}_2 \subset \mathcal{B}}{\Big|C^L(\mathcal{B}_1) - C^L(\mathcal{B}_2)\Big|}, 
\end{align}
where $\mathcal{B}_1$ and $\mathcal{B}_2$ are different subsets of the belief set. $C^S(\mathcal{B}_1)$ and $C^L(\mathcal{B}_1)$ are the incurred cost of the speaker for transmitting $\mathcal{B}_1$ and incurred cost of the listener for inferring $\mathcal{B}_2$, respectively.
The denominator of $D_1$ captures the summation of the probability of transiting to the initial and final event from event $e_j$. The denominator of $D_2$ implies that, by going forward in a TEC, the probability of transiting to the final event must increase.
\end{thm}
\vspace{-5mm}
\begin{IEEEproof} See Appendix \ref{Theorem_I_Proof}. 
\end{IEEEproof}
\vspace{-1mm}
Theorem \ref{thm1}'s condition states that the gained reward of executing a task by using a perfect description should be greater than the additional incurred cost for using that perfect description compared to imperfect descriptions ($\max(\Delta^S, \Delta^L)$). To guarantee \eqref{Equation_Constraint_Total}, the speaker and listener must know $\Delta^S$, $\Delta^L$, and the characteristics of matrices $\boldsymbol{\Tilde{P}}$ and $\boldsymbol{\Tilde{P}}$ to determine the denominators of \eqref{Equation_Constraint_Inside_Final} and \eqref{Equation_Constraint_Outside_Final}.
In practice, the speaker and listener can use their initial interaction with the environment to compute these parameters and satisfy the condition. In particular, the interaction with the environment gives the speaker and listener insight into the parameters that contribute to $D_1$ and $D_2$. Then, they can estimate an upper bound for $D_1$ and $D_2$, thereby satisfying \eqref{Equation_Constraint_Total}.
Theorem \ref{thm1}'s condition can also help improve
the convergence rate of the multi-agent RL problem of each CL step. Without meeting \eqref{Equation_Constraint_Total}, the multi-agent RL problem of CL steps can still determine the possible perfect descriptors of each event. However, the convergence time will be larger than the case in which \eqref{Equation_Constraint_Total} is satisfied.
\vspace{-5mm}
\subsection{Complexity Analysis}
\vspace{-2mm}
We next compare the total complexity of the CL method with classical RL by comparing the product of the sizes of the state and action spaces size for both methods. This product reflects the number of $Q$ values. In classical RL, the size of the state space is equal to the number of events, $|\mathcal{E}|$, and the size of the action space is equal to the total number of possible descriptions, $2^B$, where $B$ is the total number of the beliefs. The total number of $Q$ values in classical RL will be $|\mathcal{E}| \times 2^B$, and the complexity of classical RL will be $O(|\mathcal{E}| \times 2^B)$. For CL, we have $K-1$ steps, where $K$ is the number of hierarchy levels. In each CL step, we have a multi-agent RL problem including the speaker and listener as the agents. The total size of the state space is equal to the product of the speaker and listener state space sizes. The total size of the action space is equal to the product of the speaker and listener action space sizes. Thus, in the first CL step, the total size of the state and action spaces will be $|\mathcal{E}| \times B$ and $B \times B-1$, respectively. The complexity of the first step of the CL method will be $O(|\mathcal{E}| \times B^3)$. For CL step $l$, the total size of the state space is $|\mathcal{E}| \times |\Omega_{\Lambda, z_l}^L|$, where $|\Omega_{\Lambda, z_l}^L|$ is the size of the listener's state space. In CL step $l$, the total size of the action space is $|\Omega_{A, z_l}^S| \times |\Omega_{A, z_l}^L|$, where $|\Omega_{A, z_l}^S|$ and $|\Omega_{A, z_l}^L|$ are the sizes of the speaker and listener action spaces. According to \eqref{Equation:CL_Speaker_l_State_Action} and  \eqref{Equation:CL_Listener_l_State}, the maximum value of $|\Omega_{A, z_l}^S|$ and $|\Omega_{\Lambda, z_l}^L|$ is $B \choose l$, which captures all possibilities of selecting $l$ beliefs. Also, from \eqref{Equation:CL_Listener_l_Action}, the maximum value of $|\Omega_{A, z_l}^L|$ is $B \choose 1$. Then, the maximum value of the product of the size of total state space and size of total action space at step $l$ is $|\mathcal{E}| \times {B \choose l} \times {B \choose l} \times {B \choose 1}$ and the complexity of the learning method at step $l$ is $O(|\mathcal{E}| \times B^{2l+1})$. We thus observe that the complexity of the learning in each step of the CL method is a polynomial function of $B$. However, the complexity of the classical RL algorithm, $O(|\mathcal{E}| \times 2^B)$, is an exponential function of $B$. Clearly, the proposed CL method has a significantly lower complexity compared to classical RL. 
\vspace{-6mm}
\section{Simulation results and analysis}
\vspace{-2.2mm}
\label{Section:Numerical_Results}
For our simulations, we consider a hierarchical belief set with $22$ beliefs divided across $4$ levels with $4$ beliefs in level 1, $5$ beliefs in level 2, $6$ beliefs in level 3, and $7$ beliefs in level 4. The number of beliefs increases with the hierarchy level because higher hierarchy levels have more semantic information.
Also, the transmission and inference costs of the beliefs increase by going deeper in the hierarchy. We randomly generate the transmission cost of the beliefs in the first level with values between $1$ and $2$, the beliefs in the second level with values between $2$ and $3$, and we proceed similarly for the remaining levels. The listener's inference cost for each belief is assumed to be half of the transmission cost of that belief.
We consider $120$ events, including initial, intermediary, and final events. The perfect description of each event can include $2$ to $4$ beliefs. We consider $30$ tasks and assume the minimum and maximum number of events in each task to be $3$ and $6$. Matrices $\boldsymbol{P}$ and $\boldsymbol{\tilde{P}}$ are of size $120 \times 120$, and they are generated according to the TEC of each task.
For comparison, we use a classical RL solution based on Q-learning in which, to solve \eqref{Equation:Optimization}, the speaker explores different subsets of the belief set using the $\epsilon$-greedy policy and without the cooperation of the listener and also a CL without inference scheme.
\begin{figure}
     \centering
     \includegraphics[width=8cm]{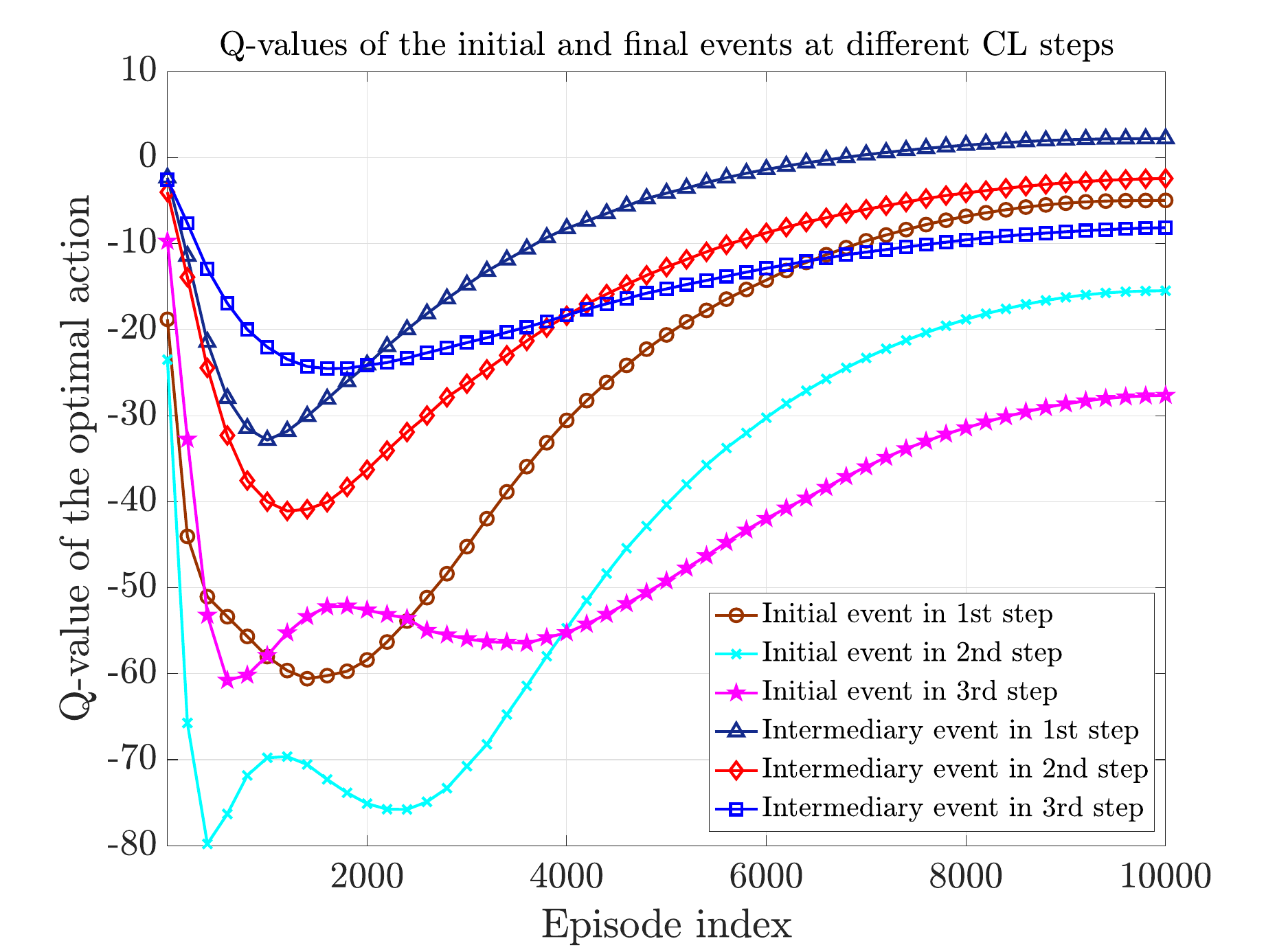}
	 \vspace{-6mm}
	\caption{\small{Comparison of the convergence of the CL method for initial and intermediary events.}}
	\label{Fig_Convergence} 
	\vspace{-8mm}
\end{figure}

\begin{figure}[!t]
\centering
\vspace{-1mm}
	\includegraphics[width=8cm]{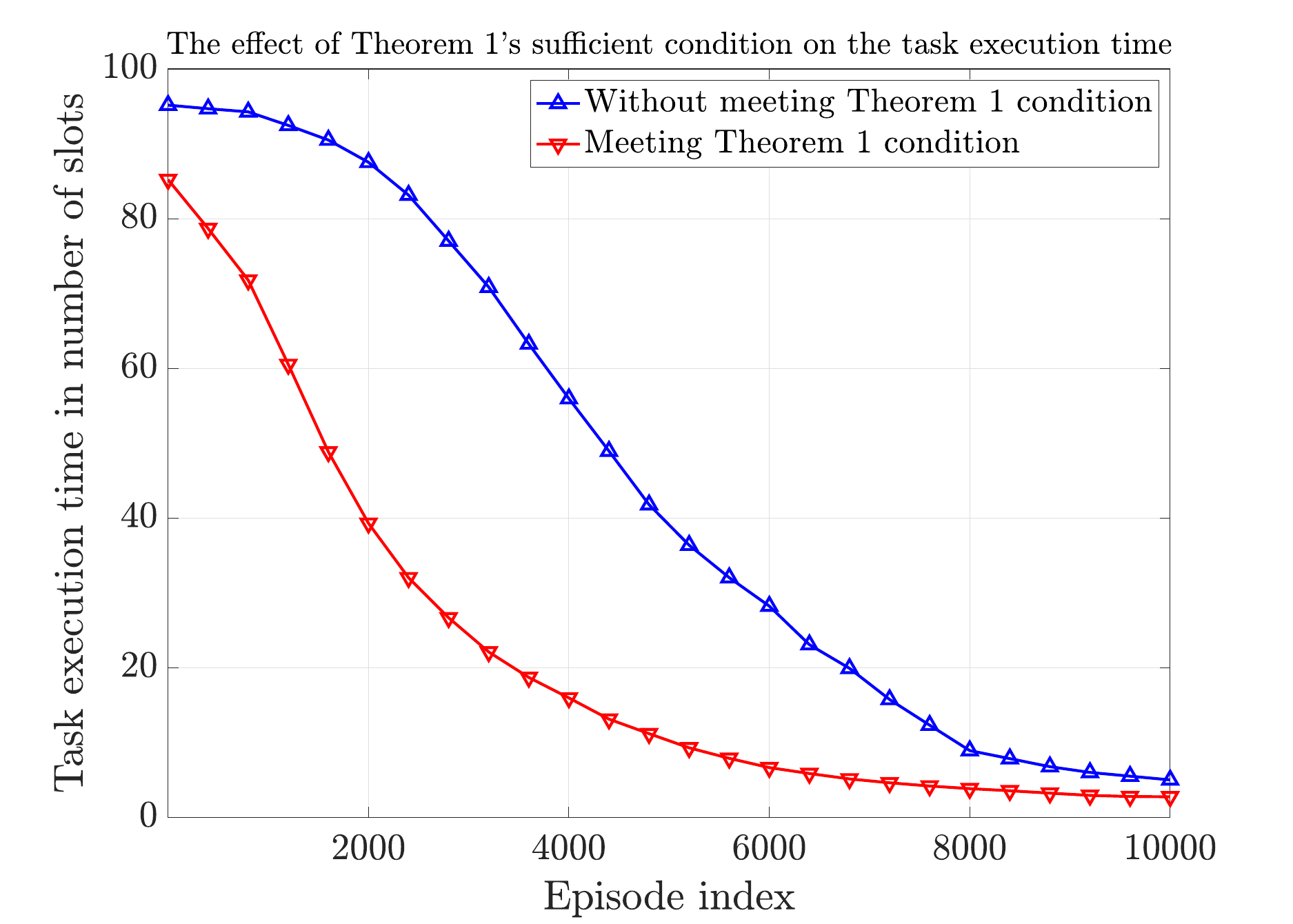} 
	\vspace{-6mm}
	\caption{\small{The effect of Theorem $1$ on the performance of the proposed CL method.}}
	\label{Fig_SufficientCondition} 
	\vspace{-10mm}
\end{figure}
\vspace{-6mm}
\subsection{Convergence of the Proposed CL Method}
\vspace{-2mm}
For investigating the convergence of the CL method, we can separately evaluate the convergence of the CL's steps. In Fig. \ref{Fig_Convergence}, the $Q$-values of the optimal action for initial and intermediary events at the speaker side and for different steps of the CL method are presented. Fig. \ref{Fig_Convergence} shows that the multi-agent RL problem of each step converges in $10,000$ episodes. The required number of episodes depends on the total number of beliefs, events, and tasks. 
Fig. \ref{Fig_Convergence} shows that the convergence of intermediary events is faster than the convergence of initial events. This is because, in the intermediary events, by transmitting a perfect description, the system can transit to a final event and gain an immediate reward. However, by transmitting a perfect description in the initial event, the system can only transit to an intermediary event with no immediate reward. Fig. \ref{Fig_Convergence} also shows a decrease in the $Q$-values at the beginning of the learning since the listener needs some episodes to find an appropriate inference for each action of the speaker. When the listener finds the appropriate inferred description, the $Q$-value of the optimal action of the speaker begins to increase until it finally converges. Fig. \ref{Fig_Convergence} shows that the convergence rate in the higher steps of the CL method is slower because of the increase in the size of the action space in the higher steps.

Now, we shed light on the result of Theorem \ref{thm1} and the effect of the sufficient condition on the performance of the system. For this purpose, Fig. \ref{Fig_SufficientCondition} evaluates the effect of the introduced constraint on the task execution time in the first CL step. Fig. \ref{Fig_SufficientCondition} includes $10,000$ episodes of the first step of the proposed CL method for both cases of meeting and not meeting the condition of Theorem \ref{thm1}. Fig. \ref{Fig_SufficientCondition} shows that, by meeting the condition in Theorem \ref{thm1}, the convergence time of the proposed algorithm will be faster, and the task execution time will be smaller. Precisely, when the system can meet the condition of Theorem \ref{thm1}, it can achieve about $45.4\%$ and $44\%$ improvement in task execution time compared to the scenario without meeting the Theorem \ref{thm1} condition during the learning and at the end of learning, respectively.

\begin{figure}
     \centering
     \begin{subfigure}[b]{0.45\textwidth}
         \centering
         \includegraphics[width=\textwidth]{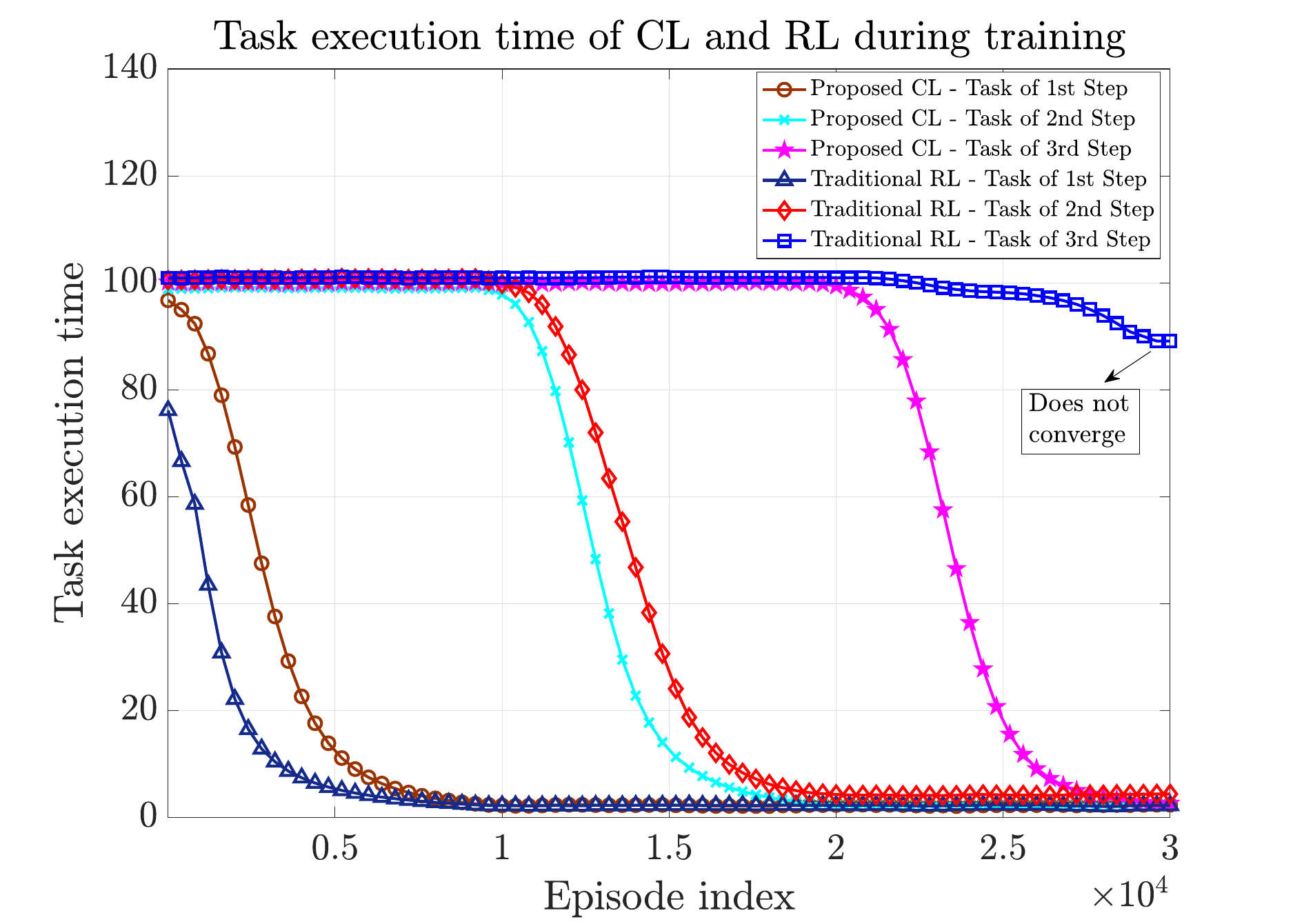}
         \vspace{-10mm}
         \caption{\scriptsize{Task execution time.}}
         \label{Fig_Training_Task_Time}
     \end{subfigure}
     \hfill
     \begin{subfigure}[b]{0.45\textwidth}
         \centering
         \includegraphics[width=\textwidth]{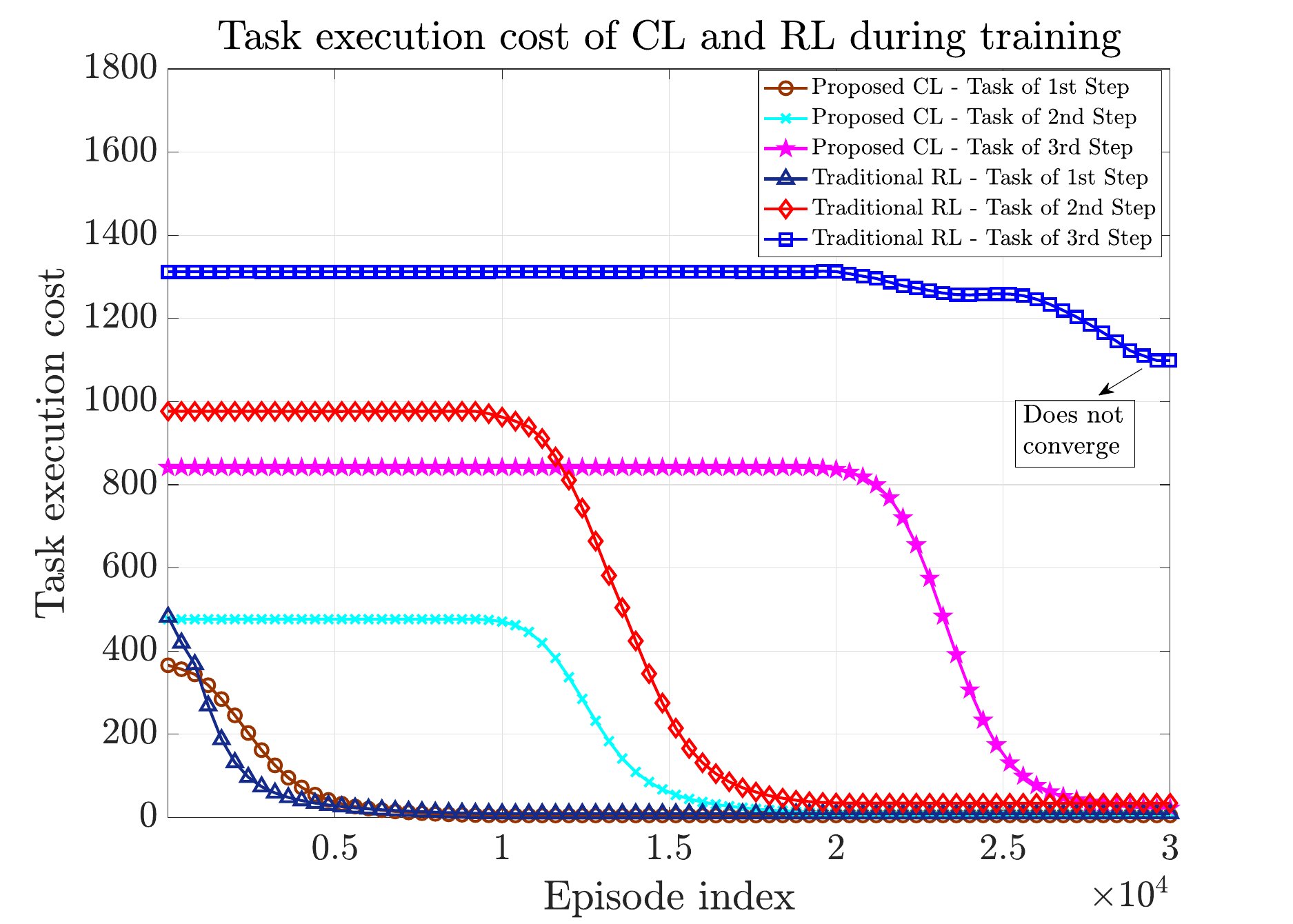}
         \vspace{-10mm}
         \caption{\scriptsize{Task execution cost.}}
         \label{Fig_Training_Task_Cost} 
     \end{subfigure}
     \vspace{-5mm}
     \caption{\small{Comparison of the CL with classical RL in terms of task execution time and cost.}}
     \label{Fig_Training_Task_Time_Cost} 
      \vspace{-11mm}
\end{figure}

\begin{figure}
     \centering
     \begin{subfigure}[b]{0.45\textwidth}
         \centering
         \includegraphics[width=\textwidth]{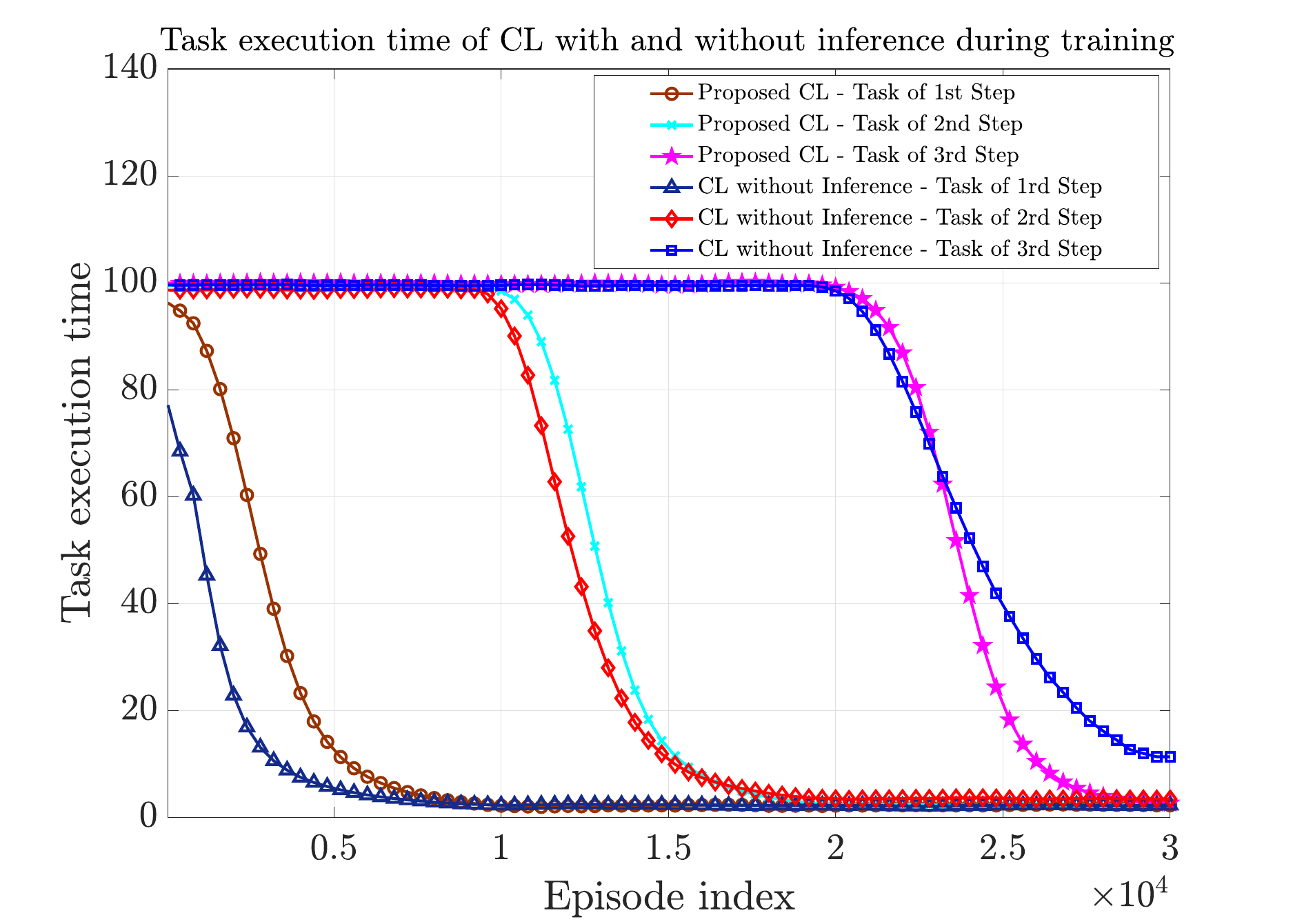}
         \vspace{-10mm}
         \caption{\scriptsize{Task execution time.}}
         \label{Fig_Training_Task_Time_Inference}
     \end{subfigure}
     \hfill
     \begin{subfigure}[b]{0.45\textwidth}
         \centering
         \includegraphics[width=\textwidth]{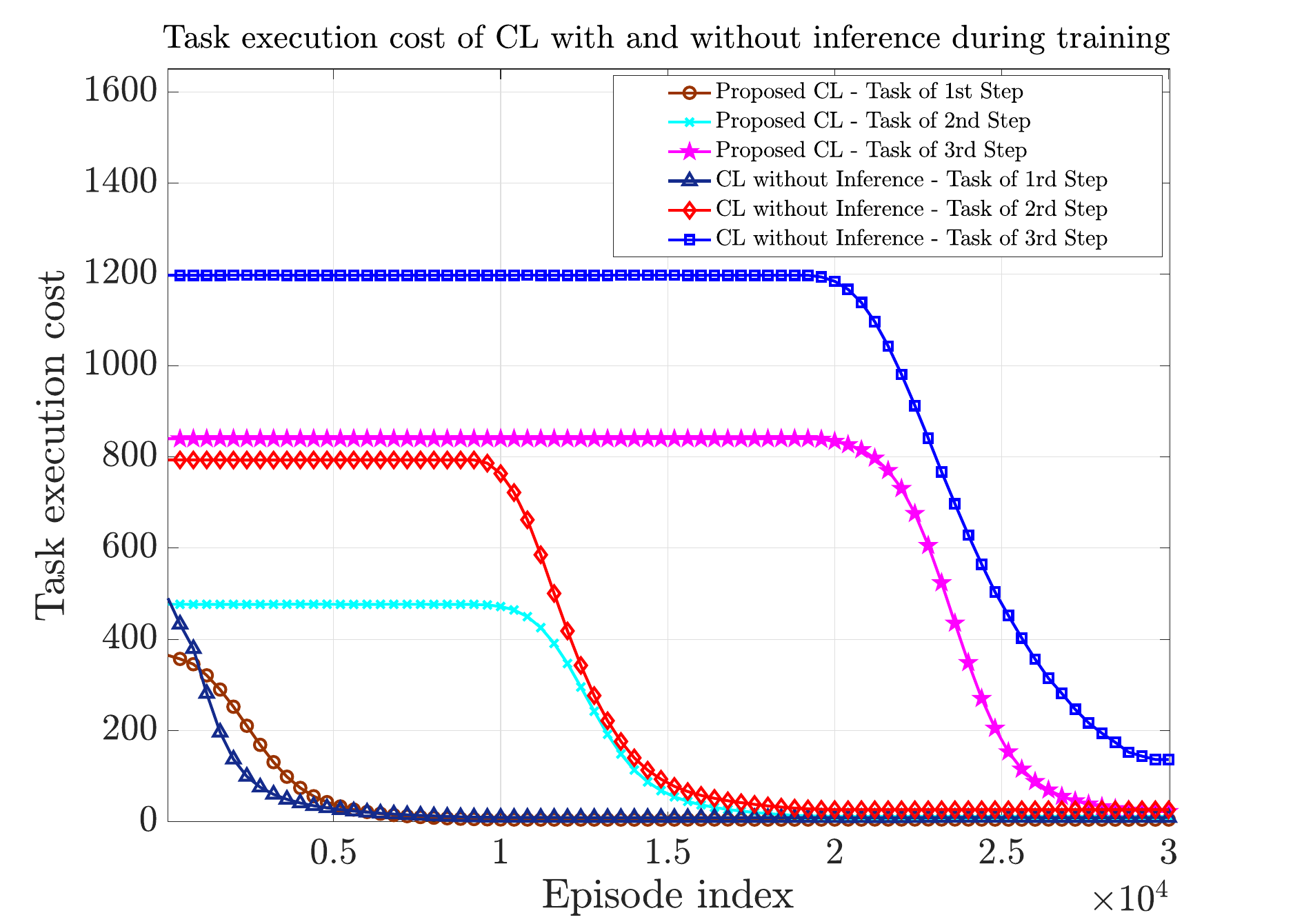}
         \vspace{-10mm}
         \caption{\scriptsize{Task execution cost.}}
         \label{Fig_Training_Task_Cost_Inference} 
     \end{subfigure}
     \vspace{-5mm}
     \caption{\small{Comparison of the proposed CL and CL without inference in terms of task execution time and cost.}}
     \label{Fig_Training_Task_Time_Cost_Inference} 
      \vspace{-12mm}
\end{figure}
\vspace{-7mm}
\subsection{Performance Comparison of Proposed CL Method and classical RL}
\vspace{-2mm}
Fig. \ref{Fig_Training_Task_Time_Cost} shows the task execution time and cost resulting from the proposed CL method and classical RL for different tasks. The tasks of step $l$ are the tasks that are executable in step $l$ of the CL method. Precisely, the TEC of the tasks of step $l$ includes events that can be described in step $l$ of the CL method.
The task execution time is equal to the length of each episode.
Figs. \ref{Fig_Training_Task_Time}-\ref{Fig_Training_Task_Cost} include $30,000$ episodes for both methods, which is a sufficient time for the convergence of the CL method. Figs. \ref{Fig_Training_Task_Time}-\ref{Fig_Training_Task_Cost} show that classical RL performs better than the CL method for the tasks of the first step, and convergence of the CL method in the first step is slower than classical RL. This is because our approach has two agents, while classical RL uses a single agent. However, for the tasks of the second and third steps, the performance of the CL method is significantly better than classical RL in terms of task execution time and cost. In particular, Figs. \ref{Fig_Training_Task_Time}-\ref{Fig_Training_Task_Cost} show that the proposed CL method can minimize the task execution time and cost in each step by determining the perfect description of each event. 
The convergence rate of classical RL is significantly reduced in the second step. Meanwhile, at the third step, classical RL cannot converge within a reasonable time.
This is because of the exponential increase in the size of the action space. Thus, the task execution time and cost of classical RL will increase significantly for the tasks of the third step. From Figs. \ref{Fig_Training_Task_Time}-\ref{Fig_Training_Task_Cost}, we observe that the proposed CL method yields $13.8\%$ and $49.7\%$ improvement compared to classical RL during the learning of the tasks of all steps, in terms of task execution time and cost, respectively.

Fig. \ref{Fig_Training_Task_Time_Cost_Inference} shows the task execution time and cost resulting from our CL method and CL without inference for different tasks, which includes $30,000$ episodes for both methods. In the CL without inference scheme, the listener does not perform inference on the transmitted description of the speaker and takes its action only based on the transmitted description of the speaker, $\mathcal{B}_{m,n}^S$. Therefore, the speaker should perform CL without cooperation of the listener. Similar to our CL scheme, the CL without inference scheme includes $K-1$ steps.
Fig. \ref{Fig_Training_Task_Time_Inference} shows that CL without inference performs better than the proposed CL scheme during learning of the tasks of the first and second steps. However, after convergence, the proposed approach is slightly superior to the CL without inference. Also, for the task of the third step, our CL approach performs significantly better than the CL without inference during learning and after convergence. Fig. \ref{Fig_Training_Task_Cost_Inference} shows that our CL method significantly outperforms the CL without inference scheme in terms of task execution cost for the task of the second and third steps.
This significant advantage of the tasks of higher steps stems from the fact that, for such tasks, the action space size becomes larger and, thus, for CL without inference, the speaker cannot individually determine the hierarchical structure of the belief set and the perfect descriptions of the events. However, the proposed CL method can efficiently find the perfect and abstract descriptions of the events in the tasks of higher steps due to cooperation between speaker and listener. Specifically, the task execution cost of our CL scheme is significantly better than CL without inference, because our approach allows the speaker to only to transmit an initial description of each event and the listener can complete the received description using its inference capability.

From Fig. \ref{Fig_Training_Task_Time_Inference}, we observe that the task execution time of the CL without inference scheme is $3.6\%$ smaller than  the proposed CL only during the learning of the tasks of all steps. Meanwhile, according to Fig. \ref{Fig_Training_Task_Cost_Inference}, the proposed CL method improves the task execution time by about $32.6\%$ compared to the CL without inference case for the tasks of all steps. Here, we note that the CL without inference scheme slightly outperforms the proposed CL method in terms of task execution time only during learning. When both methods converge, the proposed approach significantly outperforms the CL without inference scheme, as will be evident next.
\begin{figure}[!t]
\centering
\vspace{-3mm}
	\includegraphics[width = 8cm]{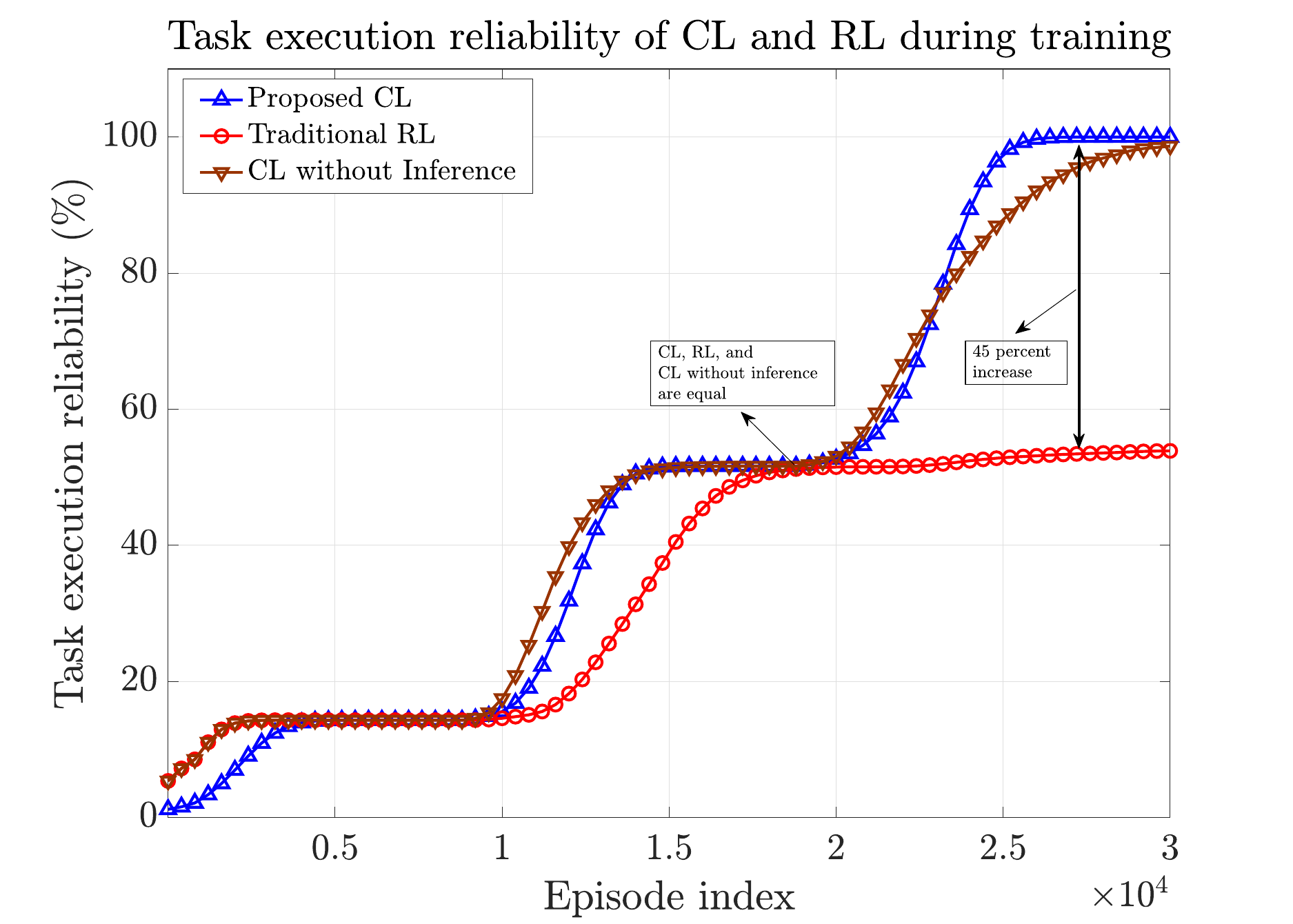}
	\vspace{-6mm}
	\caption{\small{Comparison of CL with classical RL and CL without inference in terms of task execution reliability.}}
	\label{Fig_Training_Task_Percentage} 
	\vspace{-11mm}
\end{figure}

To further highlight the advantage of our CL method, we consider a new metric called \emph{task execution reliability} that quantifies the percentage of successfully executed tasks during a limited number of time slots. In Fig. \ref{Fig_Training_Task_Percentage}, we present the task execution reliability resulting from the proposed CL method, classical RL, and CL without inference during different steps of the CL method. Fig. \ref{Fig_Training_Task_Percentage} shows that the proposed CL method can achieve $100\%$ task execution reliability when the algorithm converges and determines the perfect descriptor of all events at the end of the third step. In contrast, the CL without inference achieves $98.8\%$ task execution reliability when the algorithm converges.
Meanwhile, since classical RL does not converge in the third step, its task execution reliability does not improve during the third step. Thus, the task execution reliability of the CL method significantly outperforms the classical RL at the end of learning. The proposed CL method yields around $45\%$ improvement in task execution reliability, compared to classical RL, at the end of learning.
Note that if the classical RL algorithm is given sufficient time to converge, it may reach a reliability of $100\%$. However, the required convergence time increases exponentially by going forward in the CL steps and, thus, the classical RL approach cannot achieve maximum reliability with minimal delay as done by our approach.

\begin{figure}
     \centering
     \begin{subfigure}[b]{0.45\textwidth}
         \centering
         \includegraphics[width=\textwidth]{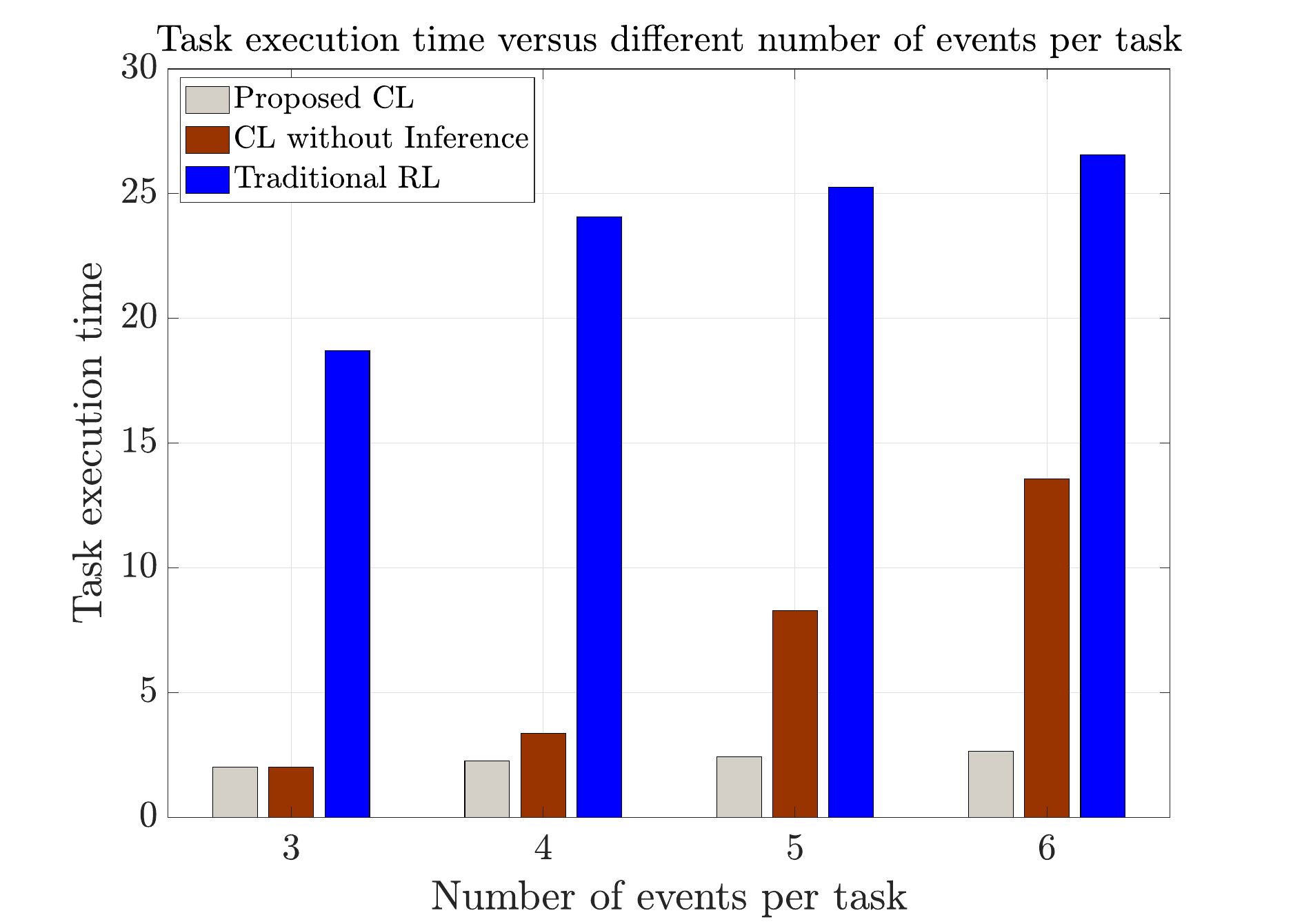}
         \vspace{-10mm}
         \caption{\scriptsize{Task execution time.}}
         \label{Fig_Final_Learning_time}
     \end{subfigure}
     \hfill
     \begin{subfigure}[b]{0.45\textwidth}
         \centering
         \includegraphics[width=\textwidth]{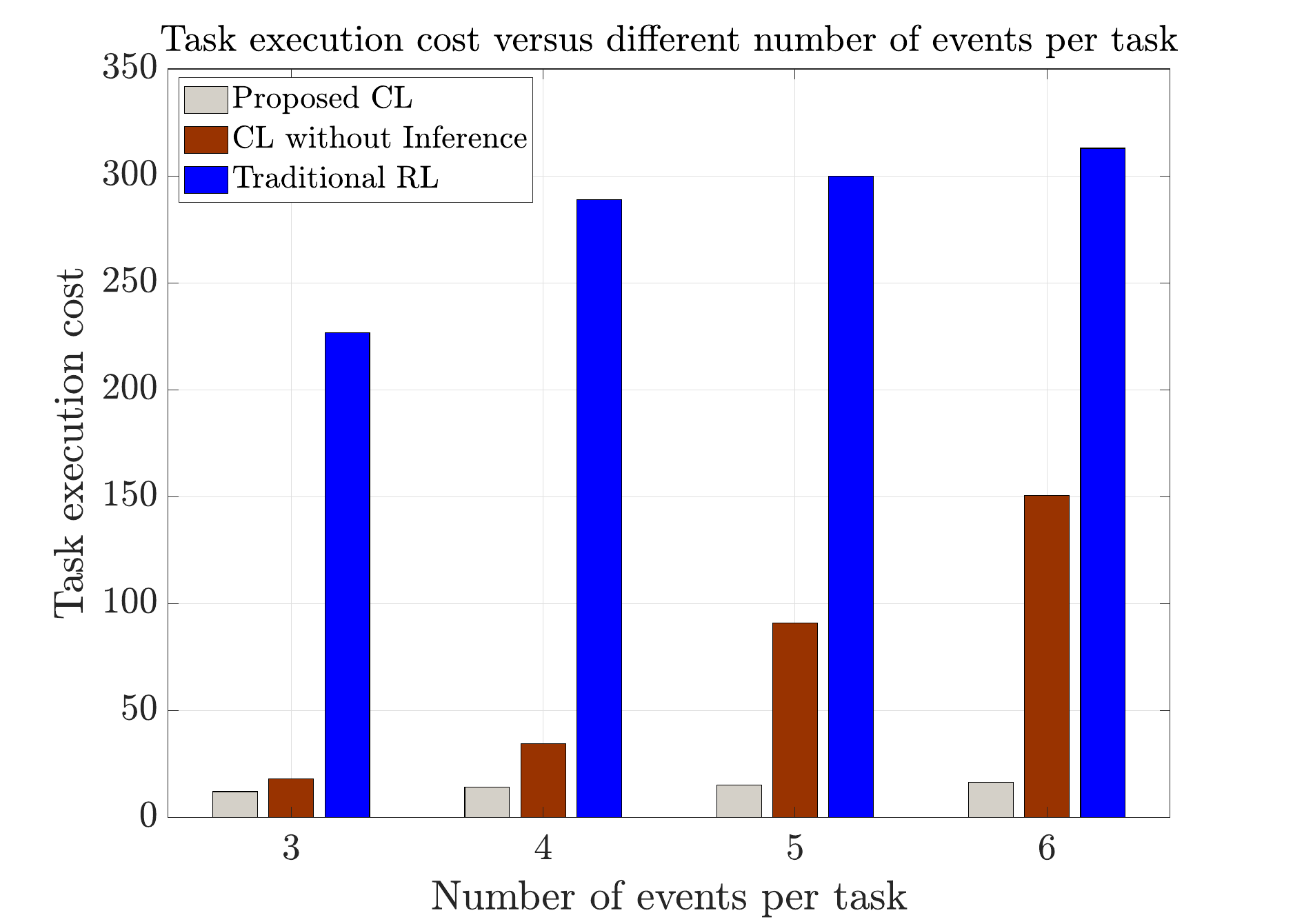}
         \vspace{-10mm}
         \caption{\scriptsize{Task execution cost.}}
         \label{Fig_Final_Learning_cost}
     \end{subfigure}
     \vspace{-5mm}
     \caption{\small{Benefit of proposed CL compared to classical RL and CL without inference for different events per task.}}
        \label{Fig_Final_Learning1} 
        \vspace{-11mm}
\end{figure}

Figs. \ref{Fig_Final_Learning_time}-\ref{Fig_Final_Learning_cost} show the task execution time and cost resulting from our CL method, classical RL, and CL without inference as a function of the number of events per task at the end of learning. Fig. \ref{Fig_Final_Learning_time} shows that the task execution time resulting from the CL method increases proportionally with the number of events. However, the task execution time resulting from classical RL is significantly higher than that of the CL method for all number of events per task because classical RL cannot converge in reasonable time to the perfect event description. Also, our CL method significantly outperforms the CL without inference in terms of task execution time, specifically as the number of events per task increases.
Fig. \ref{Fig_Final_Learning_time} shows that, the CL method yields an average of $89\%$ and $65.6\%$ improvement in task execution time compared to classical RL and CL without inference, respectively. In Fig. \ref{Fig_Final_Learning_cost}, we can see a similar improvement in task execution cost. Precisely, according to Fig. \ref{Fig_Final_Learning_cost}, our CL scheme yields average improvements of $94\%$ and $80.3\%$ compared to classical RL and CL without inference, respectively.

\begin{figure}
     \centering
     \begin{subfigure}[b]{0.44\textwidth}
         \centering
         \includegraphics[width=\textwidth]{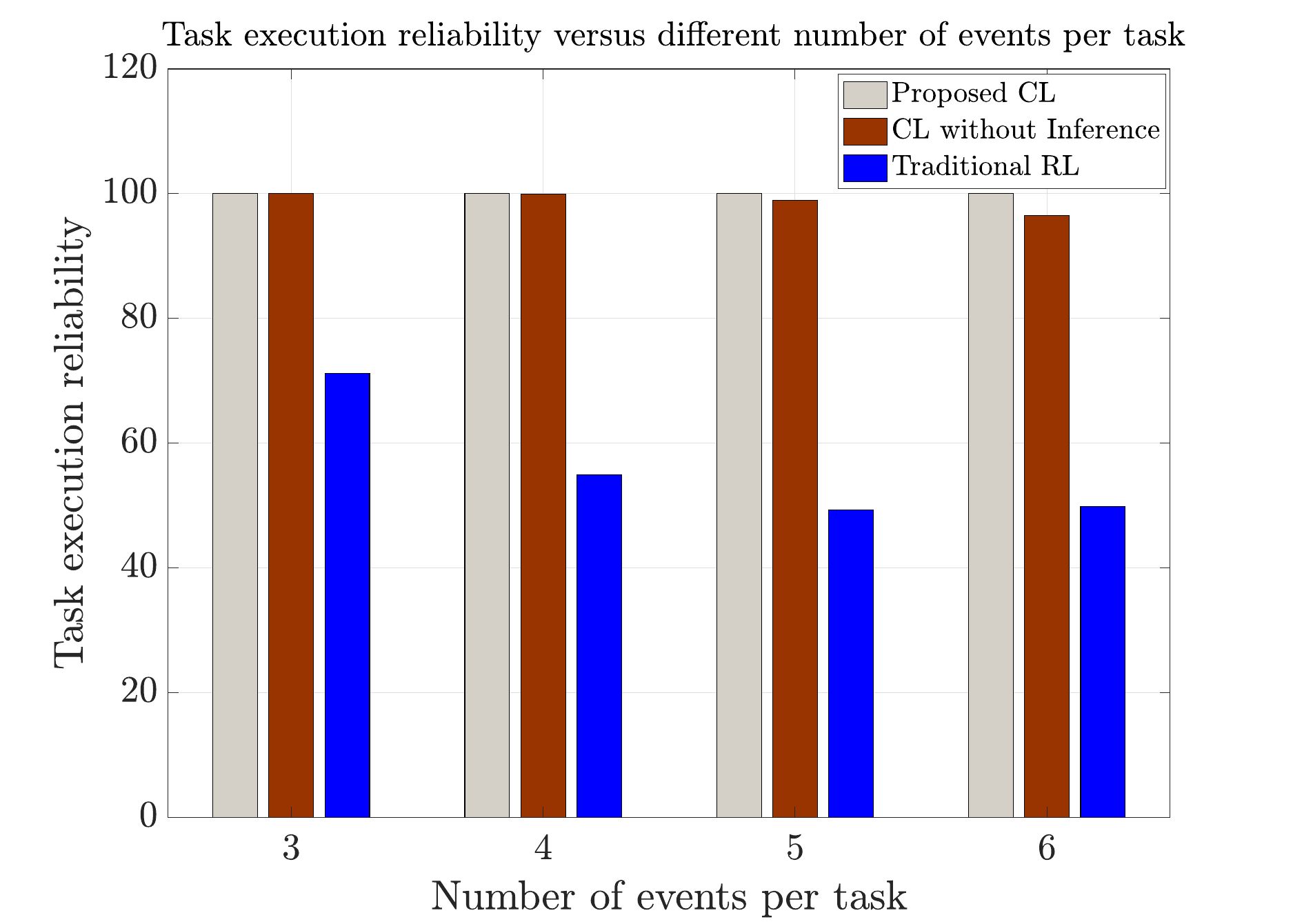}
         \vspace{-10mm}
         \caption{\scriptsize{Task execution reliability.}}
         \label{Fig_Final_Learning_Reliability}
     \end{subfigure}
     \hfill
     \begin{subfigure}[b]{0.44\textwidth}
         \centering
         \includegraphics[width=\textwidth]{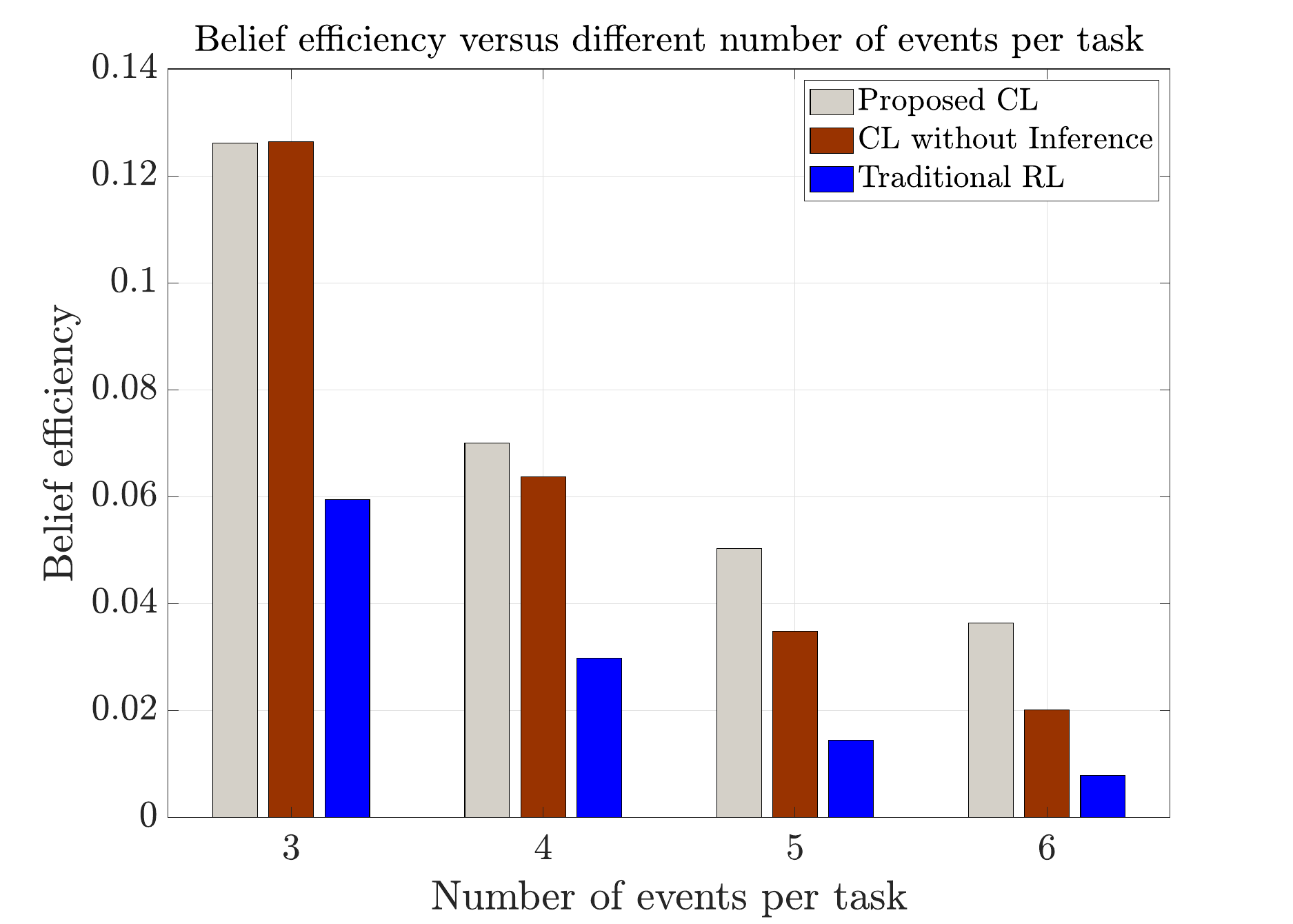}
         \vspace{-10mm}
         \caption{\scriptsize{Belief efficiency.}}
         \label{Fig_Final_Learning_Efficiency}
     \end{subfigure}
     \vspace{-5mm}
     \caption{\small{Benefit of proposed CL compared to classical RL and CL without inference for different events per task.}}
        \label{Fig_Final_Learning2} 
        \vspace{-12mm}
\end{figure}

Figs. \ref{Fig_Final_Learning_Reliability}-\ref{Fig_Final_Learning_Efficiency} show the task execution reliability and belief efficiency for the proposed CL method, classical RL, and CL without inference as a function of the number of events per task at the end of learning. Fig. \ref{Fig_Final_Learning_Reliability} shows that the task execution reliability of the CL method is approximately independent of the number of events per task. However, the task execution reliability of classical RL decreases as the number of events per task increases. This is because, when the number of events per task increases, determining the perfect description becomes more challenging. From Fig. \ref{Fig_Final_Learning_Reliability}, we observe that the proposed CL method yields, on average, about $45\%$ improvement in task execution reliability compared to classical RL. Also, Fig. \ref{Fig_Final_Learning_Reliability} shows that both the proposed CL method and CL without inference perform similarly in terms of task execution reliability.
Fig. \ref{Fig_Final_Learning_Efficiency} shows that the CL method outperforms the classical RL and CL without inference in belief efficiency for all task sizes because, in the CL method, the speaker and listener jointly determine the perfect description of each event. Fig. \ref{Fig_Final_Learning_Efficiency} shows that when the number of events per task increases, the belief efficiency will decrease because the system will require more beliefs for task execution. However, the superiority of the CL method is maintained for different task sizes. Particularly, 
Fig. \ref{Fig_Final_Learning_Efficiency} shows that the proposed CL method yields, on average, around $2.5$-fold and $15.5\%$ improvement in belief efficiency compared to the classical RL and CL without inference, respectively.

\vspace{-7mm}
\subsection{Channel Error Effect on Proposed CL Method}
\vspace{-2mm}
In this work, we assumed that the wireless channel between the speaker and listener is perfect and, thus, the listener can receive the transmitted description of the speaker without error. This is a common assumption in the semantic communication literature \cite{seo2021semantics}. However, it is useful to investigate the channel errors. In order to take into account the effect of wireless channel error in the proposed CL method, we can consider the existence of an error probability in the transmission of each belief, named belief transmission error and indicated by $p(\text{error})$. The value of $p(\text{error})$ indicates the probability that an error occurs in a transmitted belief from the speaker to the listener. For simplicity, we assumed that all beliefs of the belief set have an equal error probability. Extension to the case in which each belief has a specific error probability is straightforward. For simplicity, we assume that even if only one of the beliefs in the description $\mathcal{B}_{m,n}^S$ experiences error during transmission, the listener cannot utilize that description.

\begin{figure}
     \centering
     \begin{subfigure}[b]{0.45\textwidth}
         \centering
         \includegraphics[width=\textwidth]{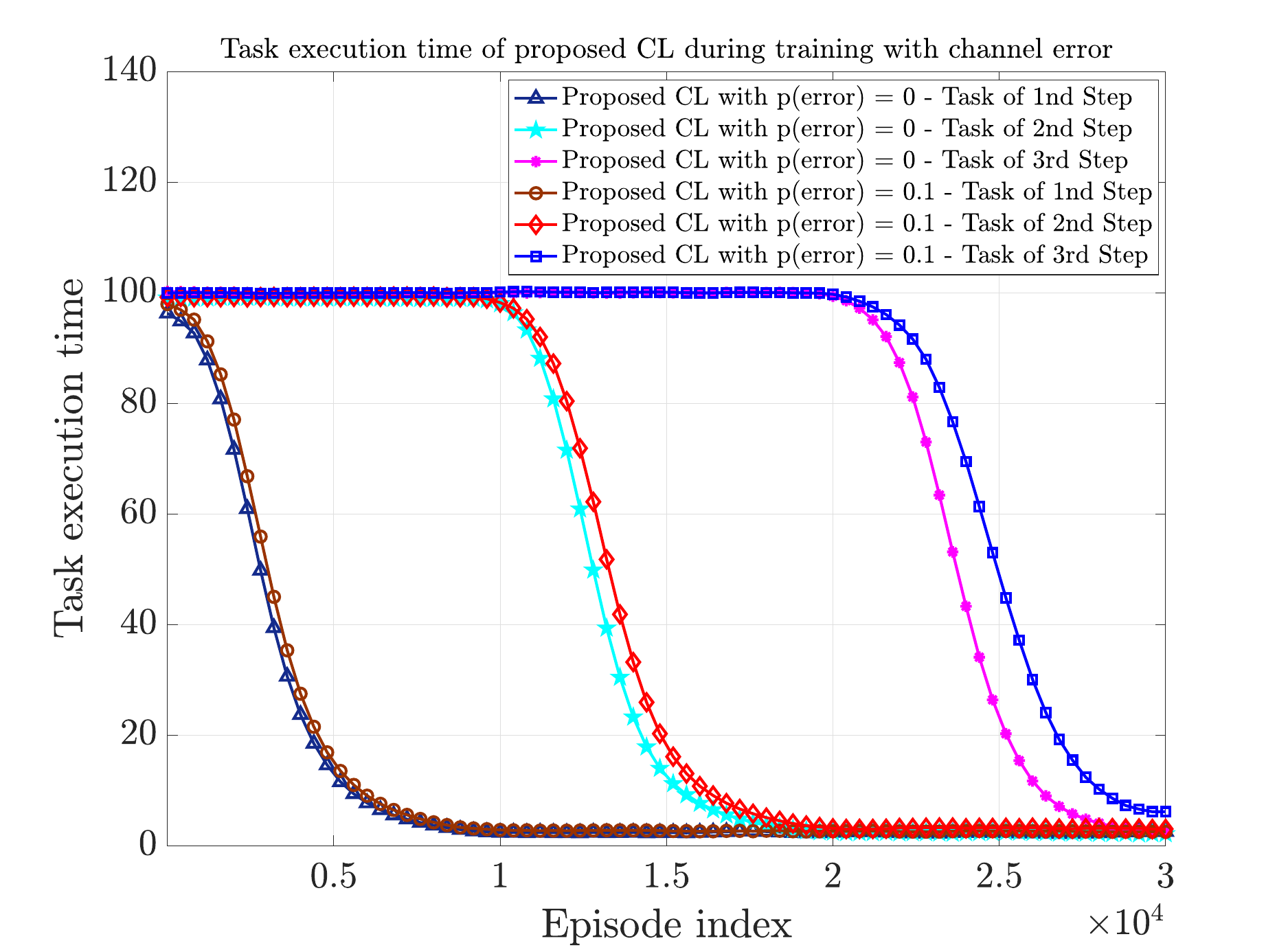}
         \vspace{-10mm}
         \caption{\scriptsize{Task execution time.}}
         \label{Fig_Channel_Error_Time}
     \end{subfigure}
     \hfill
     \begin{subfigure}[b]{0.45\textwidth}
         \centering
         \includegraphics[width=\textwidth]{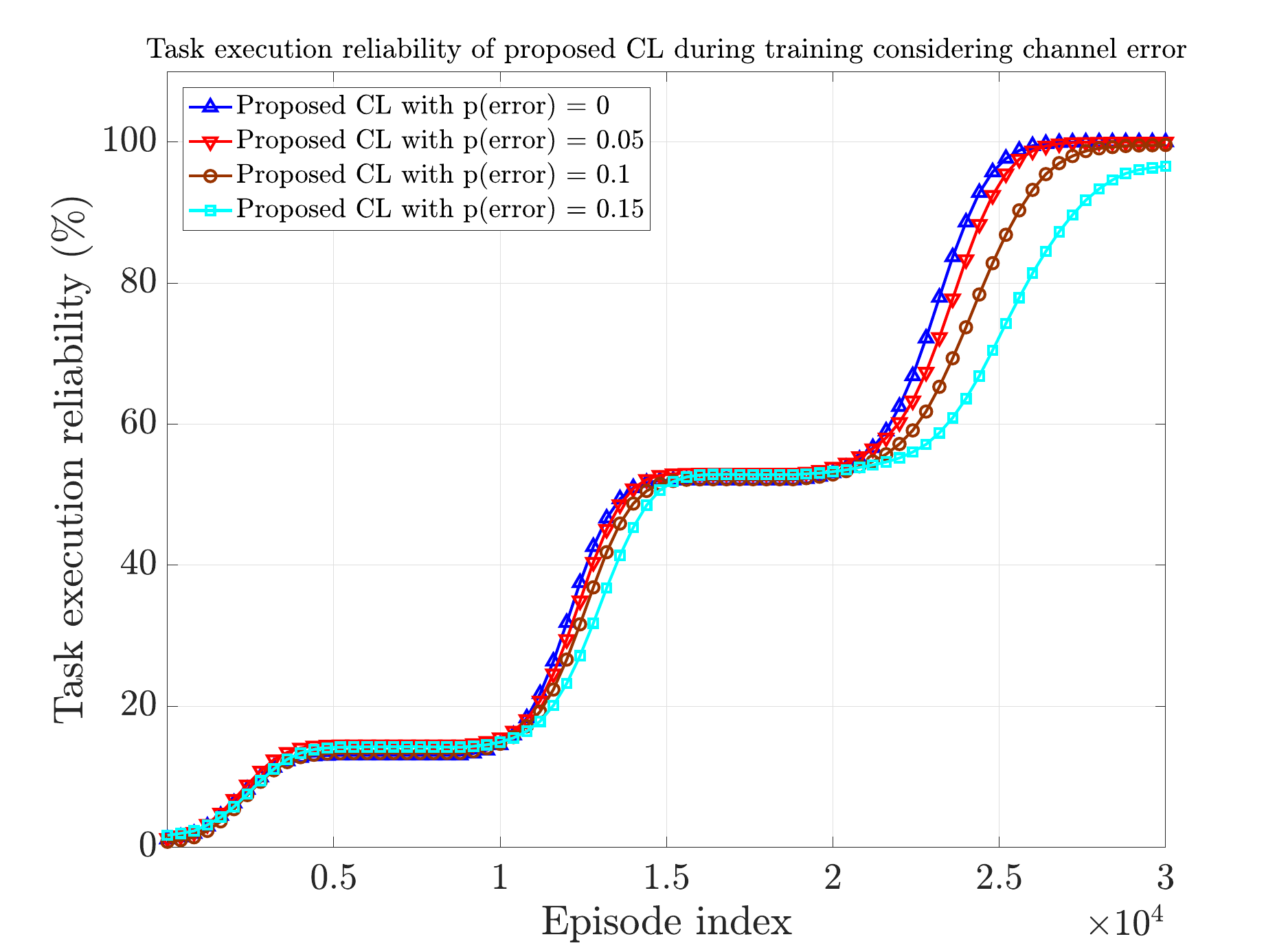}
         \vspace{-10mm}
         \caption{\scriptsize{Task execution reliability.}}
         \label{Fig_Channel_Error_Reliability}
     \end{subfigure}
     \vspace{-6mm}
     \caption{\small{Effect of wireless channel error on the performance of the proposed CL method.}}
        \label{Fig_Channel_Error}
        \vspace{-12mm}
\end{figure}

Figs. \ref{Fig_Channel_Error_Time}-\ref{Fig_Channel_Error_Reliability} show the effect of different values of belief transmission error in the proposed CL method, in terms of task execution time and reliability during the learning, respectively. Precisely, Fig. \ref{Fig_Channel_Error_Time} shows the task execution time of the proposed CL method in two cases of $p(\text{error}) = 0$ and $p(\text{error}) = 0.1$, respectively. We know that $p(\text{error}) = 0.1$ is a relatively large value for error probability of each transmitted belief. However, for such a large value of $p(\text{error})$, the performance of the proposed CL method is only slightly affected. We can observe that only for the task of the third step which requires more beliefs for their event description, the performance of the proposed CL is slightly degraded.
According to Fig. \ref{Fig_Channel_Error_Time}, considering $p(\text{error}) = 0.1$ for each transmitted belief increases the task execution time of the task of all steps by only $5.11\%$ compared to $p(\text{error}) = 0$.
Fig. \ref{Fig_Channel_Error_Reliability} shows the task execution reliability for different values of $p(\text{error})$. According to Fig. \ref{Fig_Channel_Error_Reliability}, for $p(\text{error}) = \{0, 0.05, 0.1\}$, the proposed CL method reaches $100\%$ task execution reliability and for $p(\text{error}) = 0.15$, the proposed CL method reaches $96.6\%$ task execution reliability. We can observe that for such a large value of $p(\text{error})$, the performance of the proposed CL method is only slightly degraded.

\vspace{-8mm}
\section{Conclusion}
\vspace{-3mm}
\label{Section:Conclusion}
In this paper, we have investigated the problem of semantic communication for goal-oriented networks. We have introduced a novel model that allows a speaker and listener to jointly execute system tasks based on observed environmental events. In the introduced model, the speaker sends an abstract description of its own observation of the events. Then, the listener infers and completes the transmitted description of the speaker.
For the communication between the speaker and listener, we have considered the existence of a common language consisting of a hierarchical belief structure that is based on the amount of semantic information that can be conveyed and the data type. Then, we have introduced an optimization problem to find the abstract and perfect description of each event to minimize the task execution cost with constraints on the task execution time and belief efficiency.
To solve the introduced optimization problem, we have developed a novel CL framework that can gradually identify the structure of the hierarchical belief set and the perfect description of each event using the identified portion of the belief set. Simulation results show that our CL solution significantly outperforms the classical RL and CL without inference scheme in terms of task execution time, cost, reliability, and belief efficiency.
\vspace{-14mm}
\appendix
\vspace{-4mm}
\label{Appendix_Theorem_1}
\vspace{-2mm}
\subsection{Proof of Theorem \ref{thm1}} \label{Theorem_I_Proof}
To solve the multi-agent RL problem at each CL step, we use the agent-based decomposition method from Q-learning \cite{kok2006collaborative}. In this method, for each state-action pair of each agent, we have a local $Q-$function that we update with interaction with the environment. For example, in the first CL step, for each action-state pair $(e_j, b_{i_1})$ of the speaker, where $e_j \in \Omega_{\Lambda,z_1}^S$ and $b_{i_1} \in \Omega_{A,z_1}^S$, we have a local $Q-$function $Q^S(e_j, b_{i_1})$, updated as follows:
\vspace{-3mm}
\begin{align}
    Q^S(e_j,b_{i_1}) := Q^S(e_j,b_{i_1}) + \beta \cdot \Big[R^S(e_j,b_{i_1}) + \gamma \cdot \max_{b_{i_1^{\prime}}}{Q^S(e_{j^{\prime}},b_{i_1^{\prime}})} - Q^S(e_j,b_{i_1})\Big],
    \label{Equation_Q_Update_S}
\end{align}
where $\beta$ is the learning rate, $R^S(e_j,b_{i_1})$ is the gained reward of the speaker for taking action $b_{i_1}$ in state $e_j$, and $e_{j^{\prime}}$ is the speaker's next state. For each pair of state and action $(b_{i_1}, b_{i_2})$ of the listener, where $b_{i_1} \in \Omega_{\Lambda,z_1}^L$ and $b_{i_2} \in \Omega_{A,z_1}^L$, the local $Q-$function $Q^L(b_{i_1}, b_{i_2})$, updated as:
\vspace{-3mm}
\begin{align}
    Q^L(b_{i_1}, b_{i_2}) := Q^L(b_{i_1}, b_{i_2}) + \beta \cdot \Big[R^L(b_{i_1}, b_{i_2}) + \gamma \cdot \max_{b_{i_2^{\prime}}}{Q^L(b_{i_1^{\prime}}, b_{i_2^{\prime}})} - Q^L(b_{i_1}, b_{i_2})\Big],
    \label{Equation_Q_Update_L}
\end{align}
where $R^L(b_{i_1}, b_{i_2})$ is the gained reward of the listener for taking action $b_{i_2}$ in state $b_{i_1}$, and $b_{i_1^{\prime}}$ is the next state of the listener. 
Now, consider a task $T$ with TEC $\mathcal{O}^T = \big\{\big(e^T_1, e^T_2, \ldots, e^T_{L_T^{\text{max}}-1}, e^T_{L_T^{\text{max}}} \big) \big\}$. For simplicity, we drop index $T$ and consider $\mathcal{O}^T = \big\{\big(e_1, e_2, \ldots, e_{L_T^{\text{max}}-1}, e_{L_T^{\text{max}}} \big) \big\}$.
By initializing $Q^S(e_j,b_{i_1}) = 0$ and 
$Q^L(b_{i_1}, b_{i_2}) = 0$, for all state-action pairs of the speaker and listener, according to the update formula of \eqref{Equation_Q_Update_S} and \eqref{Equation_Q_Update_L}, the updated the $Q$-values of the speaker and listener, after at least one visit of all state and action, can be written as follows:
\vspace{-3mm}
\begin{align}
	Q^S&(e_j,b_{i_1}) \label{Equation_Q_Update_S_One_Visit}
	=\\ &\begin{cases} - \beta \cdot C^S(b_{i_1}) + \beta \cdot R_T \cdot \boldsymbol{P}(j, L_T^{\text{max}}) + O(\beta^2) & \{b_{i_1}, b_{i_2}\}  \in \mathcal{B}_{e_j}^P  \text{ and } 1 \leq j \leq L_T^{\text{max}}-1,  \\ 
								      - \beta \cdot C^S(b_{i_1}) - \beta  \cdot C_T \cdot \boldsymbol{\Tilde{P}}(j,1) + O(\beta^2)  & \{b_{i_1}, b_{i_2}\} \notin \mathcal{B}_{e_j}^P \text{ and } 1 \leq j \leq L_T^{\text{max}}-1,\end{cases}   \notag
\end{align}
\vspace{-8mm}
\begin{align}
	Q^L&(b_i, b_{i^{\prime}})  \label{Equation_Q_Update_L_One_Visit}
	= \\ &\begin{cases}
								      - \beta \cdot C^L(b_{i_2}) + \beta \cdot R_T \cdot \boldsymbol{P}(j, L_T^{\text{max}}) + O(\beta^2)  & \{b_{i_1}, b_{i_2}\} \in \mathcal{B}_{e_j}^P  \text{ and } 1 \leq j \leq L_T^{\text{max}}-1, \\
								    - \beta \cdot C^L(b_{i_2}) - \beta  \cdot C_T \cdot \boldsymbol{\Tilde{P}}(j,1) +  O(\beta^2) & \{b_{i_1}, b_{i_2}\} \notin \mathcal{B}_{e_j}^P \text{ and } 1 \leq j \leq L_T^{\text{max}}-1,
								      \end{cases} \notag
\end{align}
where $C^S(b_{i_1})$ and $C^L(b_{i_2})$ are the incurred cost of the speaker for transmitting belief $b_{i_1}$ and the incurred cost of the listener for inferring belief $b_{i_2}$.
Considering \eqref{Equation_Q_Update_S_One_Visit} and \eqref{Equation_Q_Update_L_One_Visit}, the sufficient conditions to converge to the perfect descriptor of each event at the end of each CL step, are:
\vspace{-4mm}
\begin{align}
      \big[R_T \cdot \boldsymbol{P}(j, L_T^{\text{max}}) - C^S(b_{i_1}) \big] &\geq -  \big[C_T \cdot \boldsymbol{\Tilde{P}}(j,1) + C^S(b_{i_1}^{\prime})\big], \; 1 \leq j \leq L_T^{\text{max}}-1, \label{Equation_Constraint_Inside_Speaker_Start}\\
      \big[R_T \cdot \boldsymbol{P}(j, L_T^{\text{max}}) - C^L(b_{i_2}) \big] &\geq  - \big[C_T \cdot \boldsymbol{\Tilde{P}}(j,1) + C^L(b_{i_2}^{\prime})\big], \; 1 \leq j \leq L_T^{\text{max}}-1, \label{Equation_Constraint_Inside_Listener_Start}\\
     \cdot \big[R_T \cdot \boldsymbol{P}(j, L_T^{\text{max}}) - C^S(b_{i_1}) \big]&\geq  \big[R_T \cdot \boldsymbol{P}(j-1, L_T^{\text{max}}) - C^S(b_{i_1}^{\prime})\big],  \; 2 \leq j \leq L_T^{\text{max}}-1,  \label{Equation_Constraint_Outside_Speaker_Start}\\
      \big[R_T \cdot \boldsymbol{P}(j, L_T^{\text{max}}) - C^L(b_{i_2}) \big]&\geq  \big[R_T \cdot \boldsymbol{P}(j-1, L_T^{\text{max}}) - C^L(b_{i_2}^{\prime})\big],  \; 2 \leq j \leq L_T^{\text{max}}-1, \label{Equation_Constraint_Outside_Listener_Start}
\end{align}
where $\{b_{i_1}, b_{i_2}\} \in \mathcal{B}_{e_j}^P$.  $b_{i_1}^{\prime}$ and $b_{i_2}^{\prime}$ are beliefs that $\{b_{i_1}^{\prime}, b_{i_2}^{\prime}\} \notin \mathcal{B}_{e_j}^P$.
We can assume $R_T = C_T$, which is not a restricting assumption because both $R_T$ and $C_T$ are used to motivate the speaker and listener for minimizing the task execution time. Now, by simplifying \eqref{Equation_Constraint_Inside_Speaker_Start} and \eqref{Equation_Constraint_Inside_Listener_Start}, we have:
\vspace{-8mm}
\begin{align}
R_T &\geq \dfrac{C^S(b_{i_1}) - C^S(b_{i_1}^{\prime})}{\big[\boldsymbol{P}(j, L_T^{\text{max}}) + \boldsymbol{\Tilde{P}}(j,1)\big]}, \quad 
R_T \geq \dfrac{C^L(b_{i_2}) - C^L(b_{i_2}^{\prime})}{\big[\boldsymbol{P}(j, L_T^{\text{max}}) + \boldsymbol{\Tilde{P}}(j,1)\big]},
\end{align}

\sloppy By combining these two constraints, and the inequalities $C^S(b_{i_1}) - C^S(b_{i_1}^{\prime}) \leq \max_{b_{i_1}, b_{i_1}^{\prime}}{\big|C^S(b_{i_1}) - C^S(b_{i_1}^{\prime})}\big|$ and $C^L(b_{i_2}) - C^L(b_{i_2}^{\prime}) \leq \max_{b_{i_2}, b_{i_2}^{\prime}}{\big|C^L(b_{i_2}) - C^L(b_{i_2}^{\prime})}\big|$, we have:
\vspace{-7mm}
\begin{align}
R_T &\geq \dfrac{\max\Big(\max_{b_{i_1}, b_{i_1}^{\prime}}{\big|C^S(b_{i_1}) - C^S(b_{i_1}^{\prime})}\big|, \max_{b_{i_2}, b_{i_2}^{\prime}}{\big|C^L(b_{i_2}) - C^L(b_{i_2}^{\prime})}\big|\Big)}{\min_{1 \leq j \leq L_T^{\text{max}}-1}{\big[\boldsymbol{P}(j, L_T^{\text{max}}) + \boldsymbol{\Tilde{P}}(j,1)\big]}} \label{Equation_Constraint_Inside_Start},
\end{align}
where inequality \eqref{Equation_Constraint_Inside_Start} should hold for each $T \in \mathcal{T}$ and, thus, we must have $R_T \geq D_1$, where $D_1$ is defined in \eqref{Equation_Constraint_Inside_Final}. Similarly, combining the constraints in \eqref{Equation_Constraint_Outside_Speaker_Start} and \eqref{Equation_Constraint_Outside_Listener_Start}, $R_T \geq D_2$, where $D_2$ is defined in \eqref{Equation_Constraint_Outside_Final}. Therefore, the sufficient condition would be $R_T \geq \max\big(D_1, D_2\big)$.
\vspace{-7mm}
\bibliographystyle{IEEEtran}
\bibliography{ref}

\begin{thebibliography}{10}
\providecommand{\url}[1]{#1}
\csname url@samestyle\endcsname
\providecommand{\newblock}{\relax}
\providecommand{\bibinfo}[2]{#2}
\providecommand{\BIBentrySTDinterwordspacing}{\spaceskip=0pt\relax}
\providecommand{\BIBentryALTinterwordstretchfactor}{4}
\providecommand{\BIBentryALTinterwordspacing}{\spaceskip=\fontdimen2\font plus
\BIBentryALTinterwordstretchfactor\fontdimen3\font minus
  \fontdimen4\font\relax}
\providecommand{\BIBforeignlanguage}[2]{{%
\expandafter\ifx\csname l@#1\endcsname\relax
\typeout{** WARNING: IEEEtran.bst: No hyphenation pattern has been}%
\typeout{** loaded for the language `#1'. Using the pattern for}%
\typeout{** the default language instead.}%
\else
\language=\csname l@#1\endcsname
\fi
#2}}
\providecommand{\BIBdecl}{\relax}
\BIBdecl

\bibitem{popovski2020semantic}
P.~Popovski, O.~Simeone, F.~Boccardi, D.~G{\"u}nd{\"u}z, and O.~Sahin,
  ``Semantic-effectiveness filtering and control for post-5{G} wireless
  connectivity,'' \emph{Journal of the Indian Institute of Science}, vol. 100,
  no.~2, pp. 435--443, 2020.

\bibitem{saad2019vision}
W.~Saad, M.~Bennis, and M.~Chen, ``A vision of 6{G} wireless systems:
  Applications, trends, technologies, and open research problems,'' \emph{IEEE
  {N}etwork}, vol.~34, no.~3, pp. 134--142, 2019.

\bibitem{chaccour2022less}
C.~Chaccour, W.~Saad, M.~Debbah, Z.~Han, and H.~V. Poor, ``Less data, more
  knowledge: Building next generation semantic communication networks,''
  \emph{arXiv preprint arXiv:2211.14343}, 2022.

\bibitem{strinati20216g}
E.~C. Strinati and S.~Barbarossa, ``6{G} networks: Beyond {S}hannon towards
  semantic and goal-oriented communications,'' \emph{Computer Networks}, vol.
  190, p. 107930, 2021.

\bibitem{luo2022semantic}
X.~Luo, H.-H. Chen, and Q.~Guo, ``Semantic communications: Overview, open
  issues, and future research directions,'' \emph{IEEE Wireless
  Communications}, vol.~29, no.~1, pp. 210--219, Feb. 2022.

\bibitem{zhou2022cognitive}
F.~Zhou, Y.~Li, X.~Zhang, Q.~Wu, X.~Lei, and R.~Q. Hu, ``Cognitive semantic
  communication systems driven by knowledge graph,'' \emph{arXiv preprint
  arXiv:2202.11958}, 2022.

\bibitem{liu2022indirect}
J.~Liu, S.~Shao, W.~Zhang, and H.~V. Poor, ``An indirect rate-distortion
  characterization for semantic sources: General model and the case of
  {G}aussian observation,'' \emph{arXiv preprint arXiv:2201.12477}, 2022.

\bibitem{yang2022semantic}
W.~Yang, Z.~Q. Liew, W.~Y.~B. Lim, Z.~Xiong, D.~Niyato, X.~Chi, X.~Cao, and
  K.~B. Letaief, ``Semantic communication meets edge intelligence,''
  \emph{arXiv preprint arXiv:2202.06471}, 2022.

\bibitem{zhou2021semantic}
Q.~Zhou, R.~Li, Z.~Zhao, C.~Peng, and H.~Zhang, ``Semantic communication with
  adaptive universal transformer,'' \emph{IEEE Wireless Communications
  Letters}, vol.~11, no.~3, pp. 453--457, Mar. 2022.

\bibitem{farsad2018deep}
N.~Farsad, M.~Rao, and A.~Goldsmith, ``Deep learning for joint source-channel
  coding of text,'' in \emph{2018 IEEE international conference on acoustics,
  speech and signal processing (ICASSP)}, Calgary, AB, Canada, April 2018.

\bibitem{lu2021reinforcement}
K.~Lu, R.~Li, X.~Chen, Z.~Zhao, and H.~Zhang, ``Reinforcement learning-powered
  semantic communication via semantic similarity,'' \emph{arXiv preprint
  arXiv:2108.12121}, 2021.

\bibitem{du2022rethinking}
H.~Du, J.~Wang, D.~Niyato, J.~Kang, Z.~Xiong, M.~Guizani, and D.~I. Kim,
  ``Rethinking wireless communication security in semantic internet of
  things,'' \emph{arXiv preprint arXiv:2210.04474}, 2022.

\bibitem{zhang2022unified}
G.~Zhang, Q.~Hu, Z.~Qin, Y.~Cai, G.~Yu, X.~Tao, and G.~Y. Li, ``A unified
  multi-task semantic communication system for multimodal data,'' \emph{arXiv
  preprint arXiv:2209.07689}, 2022.

\bibitem{yun2021attention}
W.~J. Yun, B.~Lim, S.~Jung, Y.-C. Ko, J.~Park, J.~Kim, and M.~Bennis,
  ``Attention-based reinforcement learning for real-time uav semantic
  communication,'' in \emph{2021 17th International Symposium on Wireless
  Communication Systems (ISWCS)}.\hskip 1em plus 0.5em minus 0.4em\relax IEEE,
  Sep. 2021, pp. 1--6.

\bibitem{lotfi2021semantic}
F.~Lotfi, O.~Semiari, and W.~Saad, ``Semantic-aware collaborative deep
  reinforcement learning over wireless cellular networks,'' in \emph{Proc. of
  IEEE International Conference on Communications (ICC)}, Seoul, South Korea,
  May 2022.

\bibitem{seo2021semantics}
H.~Seo, J.~Park, M.~Bennis, and M.~Debbah, ``Semantics-native communication
  with contextual reasoning,'' \emph{arXiv preprint arXiv:2108.05681}, 2021.

\bibitem{JMLR:v21:20-212}
\BIBentryALTinterwordspacing
S.~Narvekar, B.~Peng, M.~Leonetti, J.~Sinapov, M.~E. Taylor, and P.~Stone,
  ``Curriculum learning for reinforcement learning domains: A framework and
  survey,'' \emph{Journal of Machine Learning Research}, vol.~21, no. 181, pp.
  1--50, 2020. [Online]. Available:
  \url{http://jmlr.org/papers/v21/20-212.html}
\BIBentrySTDinterwordspacing

\bibitem{farshbafan2021common}
M.~K. Farshbafan, W.~Saad, and M.~Debbah, ``Common language for goal-oriented
  semantic communications: A curriculum learning framework,'' in \emph{Proc. of
  IEEE International Conference on Communications (ICC)}, Seoul, South Korea,
  May 2022.

\bibitem{xie2021task}
H.~Xie, Z.~Qin, and G.~Y. Li, ``Task-oriented multi-user semantic
  communications for {V}{Q}{A} task,'' \emph{IEEE Wireless Communications
  Letters}, vol.~11, no.~3, pp. 553--557, Mar. 2022.

\bibitem{xie2022task}
H.~Xie, Z.~Qin, X.~Tao, and K.~B. Letaief, ``Task-oriented multi-user semantic
  communications,'' \emph{IEEE Journal on Selected Areas in Communications},
  vol.~40, no.~9, pp. 2584--2597, 2022.

\bibitem{higgins2017scan}
I.~Higgins, N.~Sonnerat, L.~Matthey, A.~Pal, C.~P. Burgess, M.~Bosnjak,
  M.~Shanahan, M.~Botvinick, D.~Hassabis, and A.~Lerchner, ``Scan: Learning
  hierarchical compositional visual concepts,'' \emph{arXiv preprint
  arXiv:1707.03389}, 2017.

\bibitem{loeliger2004introduction}
H.-A. Loeliger, ``An introduction to factor graphs,'' \emph{IEEE Signal
  Processing Magazine}, vol.~21, no.~1, pp. 28--41, 2004.

\bibitem{gunantara2018review}
N.~Gunantara, ``A review of multi-objective optimization: Methods and its
  applications,'' \emph{Cogent Engineering}, vol.~5, no.~1, p. 1502242, 2018.

\bibitem{murata1996multi}
T.~Murata, H.~Ishibuchi, and H.~Tanaka, ``Multi-objective genetic algorithm and
  its applications to flowshop scheduling,'' \emph{Computers \& industrial
  engineering}, vol.~30, no.~4, pp. 957--968, 1996.

\bibitem{sutton2018reinforcement}
R.~S. Sutton and A.~G. Barto, \emph{Reinforcement learning: An
  introduction}.\hskip 1em plus 0.5em minus 0.4em\relax MIT press, 2018.

\bibitem{kok2006collaborative}
J.~R. Kok and N.~Vlassis, ``Collaborative multiagent reinforcement learning by
  payoff propagation,'' \emph{Journal of Machine Learning Research}, vol.~7,
  pp. 1789--1828, Sep. 2006.

\end{thebibliography}
\end{document}